\documentclass[twoside]{article}
\usepackage[a4paper]{geometry} 
\usepackage{mathptmx}
\usepackage{ifthen} \newboolean{WIP} \setboolean{WIP}{false}
\ifthenelse{\boolean{WIP}}
{ 
  \newcommand{\comment}[2][?]{\textbf{(#1) #2}} 
}{ 
  \newcommand{\comment}[2][]{\relax}
}
\newcommand{\cut}[1]{\relax}

\newcommand{\squeezeAboveCaption}{\relax}
\newcommand{\squeezeBelowCaption}{\relax}

\usepackage[latin1]{inputenc}
\usepackage[T1]{fontenc}

\usepackage{RR}
\RRNo{8347}
\RRdate{August 2013}

\usepackage[numbers,square,sort&compress]{natbib}
\usepackage{xspace}
\usepackage{times} 
\usepackage{amsfonts,amsmath,amssymb} 
\usepackage[defblank]{paralist} 

\usepackage{graphicx}
\usepackage{subfigure}

\usepackage{maths}

\usepackage[obeyspaces]{url} 
\usepackage{hyperref}  

\newcommand{\TransCausalPlus}{Transactional Causal+ Consistency}

\newcommand{\SwiftCloudLong}{Swift\-Cloud\xspace}

\newcommand{\SwiftCloud}{Swift\-Cloud\xspace}
\newcommand{\SwiftSocial}{Swift\-So\-cial\xspace}
\newcommand{\SwiftFS}{Swift\-Docs\xspace}

\newcommand{\SOA}{Cloud\xspace}

\newcommand{\filename}[2]{\ensuremath{\mathsf{#1\,!\,#2}}}
\newcommand{\code}[1]{{\sf #1}}

\newcommand{\DC}[1]{\ensuremath{\text{DC}_{#1}}}
\newcommand{\GTID}[2]{\ensuremath{(#1,\DC{#2})}}
\newcommand{\OTID}[2]{\ensuremath{(#1,\text{#2})}}

\newcommand{\DStarTrans}{\ensuremath{D^{*}_T}}
\newcommand{\DTrans}{\ensuremath{D_T}}

\newcommand{\VStarScout}[1]{\ensuremath{V^{*}_\mathit{#1}}}
\newcommand{\VScout}[1]{\ensuremath{V_\mathit{#1}}}
\newcommand{\VDC}{\ensuremath{V_\mathit{DC}}}
\newcommand{\Vi}[1]{\ensuremath{V_\mathit{\DC{#1}}}}

\newcommand{\maxOTID}{\ensuremath{\mathit{maxOTID}}}
\newcommand{\maxOTIDs}[2]{\ensuremath{\mathit{maxOTID}_{#1}[\mathit{#2}]}}

\RRtitle{\SwiftCloudLong: Intégration de la réplication large échelle et
  de la tolérance aux pannes jusqu'à la machine cliente}
\RRetitle{\SwiftCloudLong: Fault-Tolerant Geo-Replication Integrated all the Way to the Client Machine}

\RRauthor{Marek Zawirski {\small UPMC-LIP6 \& INRIA}\\
  Annette Bieniusa {\small U.\ Kaiserslautern}\\
  Valter Balegas {\small U.\ Nova de Lisboa}\\
  Sérgio Duarte {\small U.\ Nova de Lisboa}\\
  Carlos Baquero {\small INESC Tec \& U.\ Minho}\\
  Marc Shapiro {\small INRIA \& UPMC-LIP6}\\
  Nuno Preguiça {\small U.\ Nova de Lisboa}\\
}

\authorhead{M.\ Zawirski, A.\ Bieniusa, V.\ Balegas, S.\ Duarte, C.\ Baquero, M.\ Shapiro, N.\ Preguiça}

\RRresume{
  Afin d'améliorer le temps de réponse et la disponibilité,
  les applications web et mobiles s'appuient fréquemment sur
  l'exécution de code et le stockage de données côté machine cliente.
  Cependant, les solutions existantes sont « bricolées » et s'intégrent
  mal avec la logique côté serveur.

  Nous présentons une approche méthodique, visant à intégrer le stockage
  côté client et côté serveur.
  Elle s'appuie sur des transactions hybrides (permettant aussi bien les
  mises à jour sans conflit que fortement cohérentes), qui accèdent aux
  données, soit du côté client soit côté serveur, et qui lisent un
  instantané causalement cohérent de façon efficace.
  Si une faute se produit dans l'infrastructure, le client peut
  changer de serveur et
  reprendre immédiatement l'exécution, tout en continuant à accéder sans
  interruption à son instantané cohérent.
  Cette approche a été mise en {\oe}uvre dans \SwiftCloud{}, le premier
  système transactionnel à proposer la géo-réplication jusqu'à la
  machine cliente.

  Quelques exemples montrent que notre modèle de
  programmation est adapée à divers domaines d'application.
  Notre évaluation expérimentale montre que l'approche \SwiftCloud
  améliore la tolérance aux fautes, et que la latence et le débit sont
  améliorés d'un ordre de grandeur, par rapport aux approches de
  géo-réplication classiques.
}

\RRabstract{
  Client-side logic and storage are increasingly used in web and mobile
  applications to improve response time and availability.
  Current approaches tend to be ad-hoc and poorly integrated with the
  server-side logic.
  We present a principled approach to integrate client- and server-side
  storage.
  We support mergeable and strongly consistent transactions that target
  either client or server replicas and provide access to
  causally-consistent snapshots efficiently.
  In the presence of infrastructure faults, a client-assisted failover
  solution allows client execution to resume immediately and seamlessly
  access consistent snapshots without waiting.
  We implement this approach in \SwiftCloud{}, the first transactional
  system to bring geo-replication all the way to the client machine.
 
  Example applications show that our programming model is useful across
  a range of application areas.
  Our experimental evaluation shows that \SwiftCloud provides better
  fault tolerance and at the same time can improve both latency and
  throughput by up to an order of magnitude,
  compared to classical geo-replication
  techniques.
}

\RRmotcle{géo-réplication, cohérence causale, cohérence transactionnelle causale+, cohérence finale, tolérance aux pannes}
\RRkeyword{geo-replication, causal consistency, transactional causal+ consistency, eventual consistency, fault tolerance}

\RCParis
\RRprojet{Regal}

\RRnote{
   This research is supported in part by 
   ANR project \href{http://concordant.lip6.fr/}{ConcoRDanT}
   (ANR-10-BLAN 0208), by the Google Europe Fellowship in
   Distributed Computing awarded to Marek Zawirski, and by Portuguese 
   FCT/MCT projects PEst-OE/EEI/UI0527/2011 
   and PTDC/EIA-EIA/108963/2008 and Phd scholarship awarded to
   Valter Balegas (SFRH/BD/87540/2012).
\comment[Marek]{Any other credits?}
 }

\begin{document}
\makeRR

\section{Introduction}	
\label{sec:introduction}


Cloud computing infrastructures support a wide range of
services, 
from social networks and games to
collaborative spaces and online shops.
Cloud platforms improve availability and latency by geo-replicating
data in several data centres (DCs) across the world \cite{rep:pan:1693,db:syn:1711,fic:syn:rep:1654,rep:syn:1661,rep:syn:1662,syn:rep:1708}.
Nevertheless, the closest DC is often still too far away for an optimal
user experience.
For instance, round-trip times to the closest Facebook DC range
from several tens to several hundreds of milliseconds, and several round
trips per operation are often necessary \cite{app:1709}.
Recall that users are annoyed when interactive latency is above 50\,ms
\cite{50ms} and increased values turn them away \cite{perf-impact}.
Furthermore, mobile clients may be completely disconnected from any DC
for an unpredictable period of minutes, hours or days.



Caching data at client machines can improve latency and availability for 
many applications, and even allow for a temporary disconnection. 
While increasingly used, this approach often leads to ad-hoc 
implementations that integrate poorly with server-side storage and tend to 
degrade data consistency guarantees. 
To address this issue, we present \SwiftCloud, the first system to bring
geo-replication all the way to the client machine and to propose a
principled approach to access data replicas at client machines and 
cloud servers.

%
Although extending geo-replication to the client machine seems natural,
it raises two big challenges.
The first one is to provide programming guarantees for applications running on
client machines, at a reasonable cost at scale and under churn. 
Recent DC-centric storage systems
\cite{rep:syn:1662,syn:rep:1708,rep:syn:1661} provide transactions, and
combine support for causal consistency with mergeable objects
\cite{syn:rep:sh143}.
Extending these guarantees to the clients is problematic for a number of
reasons: standard approaches to support causality in client nodes require
vector clocks entries proportional to the number of replicas; seamless access to
client and server replicas require careful maintenance of object versions;
fast execution in the client requires asynchronous commit. 
We developed protocols that efficiently address these issues 
despite failures, by combining 
a set of novel techniques.

Client-side execution is not always beneficial.
For instance, computations that access a lot of data, such as search or
recommendations, or running strongly consistent transactions, is best done in
the DC\@.
\SwiftCloud supports server-side execution, without
breaking the guarantees of client-side in-cache execution.

The second challenge is to maintain these guarantees when the client-DC
connection breaks.
Upon reconnection, possibly to a different DC, the outcome of the
client's in-flight transactions is unknown, and state of the DC might
miss the causal dependencies of the client.
Previous cloud storage systems either retract consistency guarantees in
similar cases \cite{rep:syn:1662,syn:rep:1708,rep:syn:1690}, or avoid
the issue by waiting for writes to finish at a quorum of servers
\cite{rep:syn:1661}, which incurs high latency and may affect availability.

\SwiftCloud provides a novel client-assisted failover protocol that
preserves causality cheaply.
The insight is that, in addition to its own updates, a client may
observe a causally-consistent view of stable (i.e., stored at multiple
servers) updates from other users. This approach ensures that
the client's updates are not delayed,
and that the client's cached state matches the new DC, since it can replay
its own updates and the others are known to the DC\@.

We evaluate our protocols experimentally, and compare them with a
classical geo-replication approach.
The experiment involves three data centres in two continents, and
hundreds of remote clients.
Under sufficient access locality, \SwiftCloud enjoys order-of-magnitude
improvements in both response time and throughput over the classical approach.
This is because, not only reads (if they hit in the cache), but also
updates commit at the client side without delay; servers only need to
store and forward updates asynchronously.
Although our fault tolerance approach delays propagation, the proportion
of stale reads remains under 1\%. 



The contributions of this paper are the following:
\begin{compactitem}
\item
  The design of a cloud storage system providing geo-replication
  up to the client nodes.
\item
  The design of scalable, efficient protocols that implement the
  \TransCausalPlus{} model,
  in a system that includes replicas in the client nodes and in the
  servers.
\item
  Fault-tolerant techniques for ensuring \TransCausalPlus{} guarantees,
  without adding latency to operations.
\item
  An application study that shows the approach is useful in practice,
  and reveals where it falls short.
\item
  An experimental evaluation of the system, with different applications and
  scenarios.
\end{compactitem}
\noindent{}

\comment[Marek]{Shall we bring back the paper outline here?}

\begin{figure}[tp]
  \centering
  \includegraphics[width=.6\columnwidth]{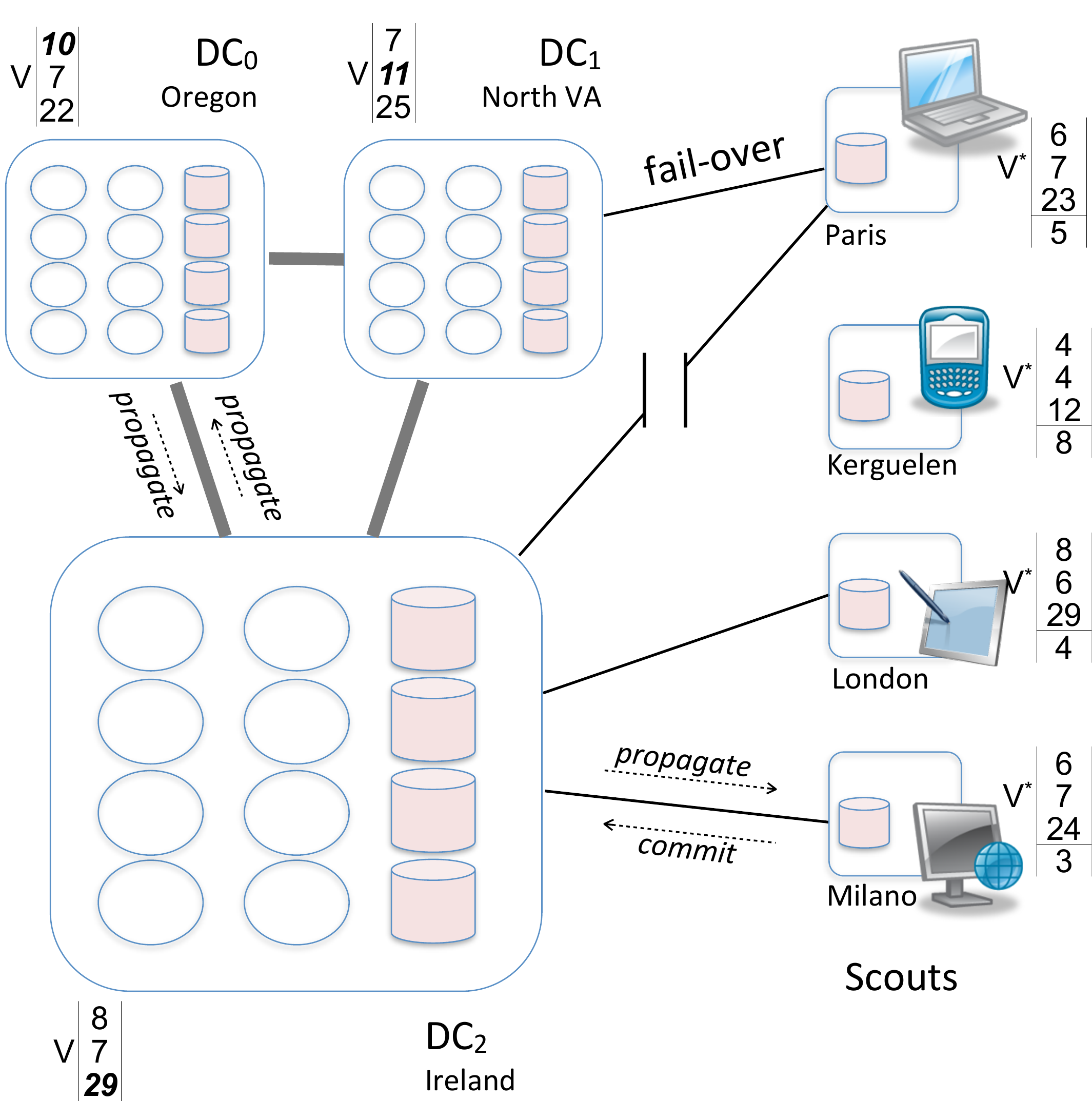}
  \squeezeAboveCaption
  \caption{\SwiftCloud system structure.}
  \squeezeBelowCaption
  \label{fig:architecture}
\end{figure}

\section{System overview}	
\label{sec:overview}


\SwiftCloud is a data storage systems for cloud platforms that spans both
client nodes and data center servers (DCs), 
as illustrated in Figure~\ref{fig:architecture}.
The core of the system consists of a set of \emph{data centres (DCs)}
that replicate every object.
At the periphery, \emph{client nodes} cache a subset of the objects. 
If the appropriate objects are in cache, responsiveness is improved and
a client node supports disconnected operation.

\subsection{Programming model}

\begin{figure}[tp]\small
\begin{verbatim}
             begin () : tx_handle
             read (tx_handle, object_id) : object
             multi_read (tx_handle, set<object_id>) : set<object>
             update (tx_handle, object, effect_op) : void
             commit (tx_handle) : void
             rollback (tx_handle) : void

             exec_stored_tx (name, params, options): set<object>
\end{verbatim}
  \squeezeAboveCaption
\caption{\label{fig:API}
  \SwiftCloud Client API.
}
\end{figure}

\SwiftCloud provides a straightforward key-object API, presented in
Figure~\ref{fig:API}.
Applications running in the client can execute sequences of read and
update operations, grouped into transactions.
Transactions can provide either strong or weak consistency, as
discussed next.

A client can request the execution of a stored transaction in the
data server.
A stored transaction is similar to a stored procedure in a database
system, and can return a set of objects.
Whereas a stored transaction runs completely in the server, a
client-side transaction contacts the server only if there is a cache
miss.
We expect that common operations will execute asynchronously in the
client cache, and that stored transactions and strongly-consistent
transactions will be rare.
For example, in a social networking application, the user's wall, and
those of his friends, can be served directly in the cache, while
computing recommendations, which requires accessing a huge number of
objects, will be implemented as a stored transaction.

\begin{figure}[tp]
  \centering
  \includegraphics[width=0.85\columnwidth]{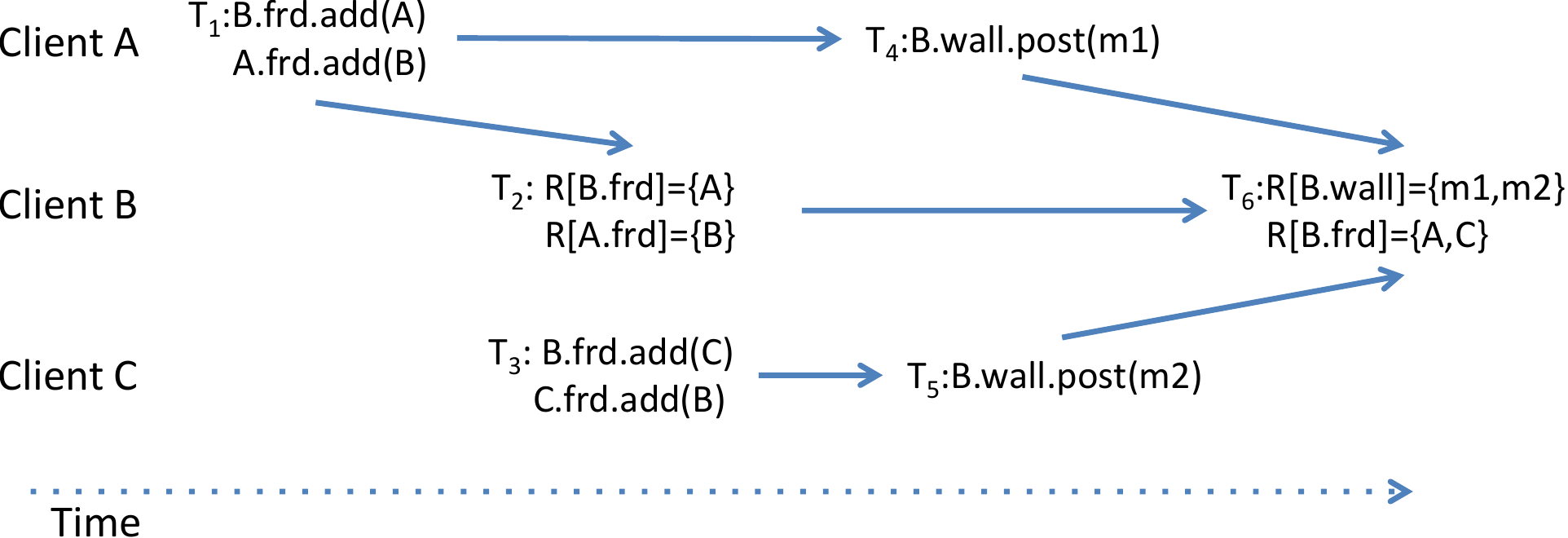}
    \squeezeAboveCaption
  \caption{Potential-Causality relation in an execution of a social networking application
(represented by arrows).}
  \squeezeBelowCaption
  \label{fig:transactions:newdeps}
\end{figure}

\subsection{Transactional model}

This section outlines our transactional model, \TransCausalPlus{}.
Intuitively, it offers the following guarantees: every transaction
reads a causally consistent snapshot; updates of a transaction are atomic
(all-or-nothing) and isolated (no concurrent transaction observes an
intermediate state); and concurrently committed updates do not conflict.

At the heart of \TransCausalPlus{} is the guarantee that
state never goes back in time: once a client has
read an object version that reflects some update, it cannot observe a
state where that update was not applied.
As an example, when user $A$ adds friend $B$ to her social network, and
later posts something, if $B$ later sees the post, she will also observe
that she is $A$'s friend.
This example is illustrated by $T_{1}$ and $T_6$ in
Figure~\ref{fig:transactions:newdeps}.
\comment[Marc]{The notation in the figure is not explained.}

Formally, we define a potential causality relation $\leadsto$ on operations
(augmenting the definition of \citet{rep:syn:1662} with transactions):
\begin{compactenum}
\item
  \emph{Execution Thread}.
  If $a$ and $b$ are two operations invoked by the same client,
  and $a$ occurs before $b$, then $a \leadsto b$.
\item
  \emph{Gets From}.
  If $u$ is an update and $r$ is a read that returns
  the value written by $u$, then $u \leadsto r$.
\item
  \emph{Transaction closure}.
\comment[Marek]{Do we bring back something like RC here? If so, the definition requires change.}
  Given $a$ and $b$, two operations of some transaction $T$, and $x$ and
  $y$, operations that are not part of $T$: if $x \leadsto a$ then $x
  \leadsto b$, and if $a \leadsto y$ then $b \leadsto y$.
\item
  \emph{Transitivity}.
  For operations $a$, $b$, and $c$, if $a \leadsto b$ and $b \leadsto
  c$, then $a \leadsto c$.
\end{compactenum}

\emph{Execution Thread} ensures that successive 
operations executed by a client are dependent. 
\emph{Gets From} ensures that a read depends on the updates it reads.
For instance, in Figure~\ref{fig:transactions:newdeps}, the read of $B.\mathit{frd}$ (the friend set of $B$)
in $T_{6}$ depends on the update to $B.\mathit{frd}$ in $T_3$: 
$B.\mathit{frd}.\mathit{add}(A) \leadsto R[B.\mathit{frd}]={A}$.
\emph{Transaction Closure} ensures transaction isolation by
extending dependence across all operations in the same transaction.
For instance, in Figure~\ref{fig:transactions:newdeps}, $B.\mathit{frd}.\mathit{add}(A) \leadsto R[B.\mathit{frd}]={A}$ is
extended to $A.\mathit{frd}.\mathit{add}(B) \leadsto R[B.\mathit{frd}]={A}$, guaranteeing that
$R[A.\mathit{frd}]={B}$.
The relation is transitive and acyclic, hence it is a partial order.

An execution of a system satisfies \TransCausalPlus{} if:
\begin{compactenum}
  \item \emph{Every transaction observes a valid snapshot and its own updates:} 
all reads of a transaction observe a state that includes all updates
from the same set of committed transactions, and earlier updates of its own
transaction, applied in a sequence that is a linear extension of $\leadsto$.
  
  \item \emph{Every snapshot is causally consistent:} the set of observed updates is
  transitively closed over $\leadsto$ and includes at least all updates committed
  earlier in $\leadsto$.
\end{compactenum}
\comment[MZ]{Even this definition is too much handwaving.}  
For instance, after $B$ observes she is a friend of $A$ and $C$, she
cannot observe that she is friend of $A$ only, since successive reads
depend on the read that showed $B$ as friend of both $A$ and $C$.

Note that this weak definition allows different clients to observe the same set of
concurrent updates applied in different orders, which poses a risk of yielding
different operation outcomes on different replicas or at different times.
We address this problem by disallowing non-commutative (order-dependent)
concurrent updates.
Practically, we enforce this property with two different types of transactions,
akin to the model of Walter~\cite{rep:syn:1661} or Red-Blue~\cite{rep:syn:1690}:
\begin{compactenum}
  \item \emph{Mergeable transaction} can update only objects with
  commutative operations and always commit. 
  \item \emph{Classic, non-mergeable transaction} can perform
  non-commutative operations, but among concurrent transactions with
  conflicting updates at most one can successfully commit.
\end{compactenum}

\subsubsection{Mergeable transactions}
The focus of this paper is the efficient and fault-tolerant support of mergeable
transactions, i.e., transactions with updates that commute with all other updates.
Mergeable transactions commute with each other and with non-mergeable transactions,
which allows to execute them immediately in the cache, commit asynchronously in
the background, and remain available in failure scenarios.

Read-only transactions are a simple case of mergeable transactions.
Concurrent updates are more difficult to handle and often complicated to merge,
with many existing systems relying on questionable heuristics, such as
last-writer-wins~\cite{db:rep:optim:1454,rep:syn:1662,syn:rep:1708}.
Our approach is to permit concurrent update transactions on dedicated
\emph{mergeable objects}.
Mergeable data types include last-writer-wins registers, the multivalue
registers of Dynamo \cite{app:rep:optim:1606}, the C-Set of
\citet{rep:syn:1661}, and a number of higher-level Conflict-free Replicated
Data Types (CRDT) of \citet{syn:rep:sh143,syn:sh144}.
CRDTs include a rich repertoire of high-level objects, such as
replicated counters, sets, maps, graphs, and sequences.

CRDTs encapsulate common concurrency and replication complexity and allow
to solve them at the system level once for all.
However, real application require either complex custom objects or using
multiple objects.
The former is impractical, whereas the latter raises new
issues, lacking cross-object guarantees~\cite{BloomL}.
Our transactional model introduces simple cross-object ordering guarantees
and allows to compose multiple objects in applications.
Examples in Section~\ref{sec:apps} suggest that for many applications
mergeable transaction can express the dominant part of the workload.
For stronger guarantees, non-mergeable transactions can be used. 

\subsubsection{Classic, non-mergeable transactions}

\SwiftCloud{} supports the traditional strongly-consistent transaction
model, in which non-commuting concurrent updates conflict (as determined
by an oracle on pairs of updates) and cannot both commit.
This primitive allows to implement arbitrarily difficult tasks, e.g.,
to build atomic registers and enforce strong data invariants when necessary.





\section{Algorithms for transactions}
\label{sec:algorithms}


\subsection{Non-mergeable transactions}

Non-mergeable transactions execute as stored procedures on the server side.
We implement a simple read-one{\slash}write-all protocol, using two-phase
commit to guarantee that no conflicting concurrent update has previously
committed.
The commit protocol could be replaced by Paxos Commit \cite{db:syn:1578}
for improved fault-tolerance.

\subsection{Mergeable transactions}

We present the algorithms used to implement mergeable transactions, first in the
failure-free case, and later (in Section~\ref{sec:FT}) in the presence of
failures.  
We assume a classical sequential model where a client executes a single
transaction at a time, i.e., a replica has a single thread of execution.
Applications interface to the system via a local module called
\emph{scout}; we assume for now that it connects to a single DC\@.
For client-side transactions, the scout is located in the client machine;
for stored transactions, the code and the scout both run in the DC\@.

An application issues a mergeable transaction by interactively executing a
sequence of reads and updates, and concludes the transaction with either a commit
or rollback.  Reads are served from the local scout; on a cache miss,
the scout fetches the data from the DC.
Updates execute in a local copy.
When a mergeable transaction terminates, its updates are applied to the
scout cache.
Eventually, they will also be committed at its DC\@.
The DC eventually propagates the effects to other DCs and other scouts
scouts as needed.

\subsubsection{System state}

The system ensures the invariant that every node (DC or scout) maintains
a causally-consistent set of object versions.
A DC replicates all objects (full replication).
The DC keeps several recent versions of an object, in order to serve the
versions requested by scouts on behalf of their transactions.
Old versions can be pruned, i.e., discarded, without impacting
correctness.

Each DC maintains a vector clock $\VDC$ that summarizes the
set of transactions that it has processed.
At \DC{i}, entry $\Vi{i}[i]$ counts the number of transactions that
\DC{i} committed.
Any other entry $\Vi{i}[j]$ counts the number of transactions committed
by \DC{j} that \DC{i} has processed.
$\Vi{i} \leq \Vi{j}$ means that \DC{i} processed a subset of the
transactions processed by \DC{j}.

A scout $S$ maintains a vector clock $\VStarScout{S}$ that summarises the 
transactions reflected by the most recent version of cached objects. 
$\VStarScout{S}$ includes one entry for each DC, plus an additional 
entry for the transactions locally committed at $S$. 
We denote with $\VScout{S}$ the same vector clock restricted to the entries for
DCs.

At all times, the globally-committed update transactions observed by a
scout are a subset of those known by its DC, i.e., $\VScout{S} \leq
\VDC$.
This invariant is obvious in the failure-free case; Section~\ref{sec:FT}
explains how we maintain it in the presence of failures.

\begin{figure}[tp]
  \centering
  \includegraphics[width=0.7\columnwidth]{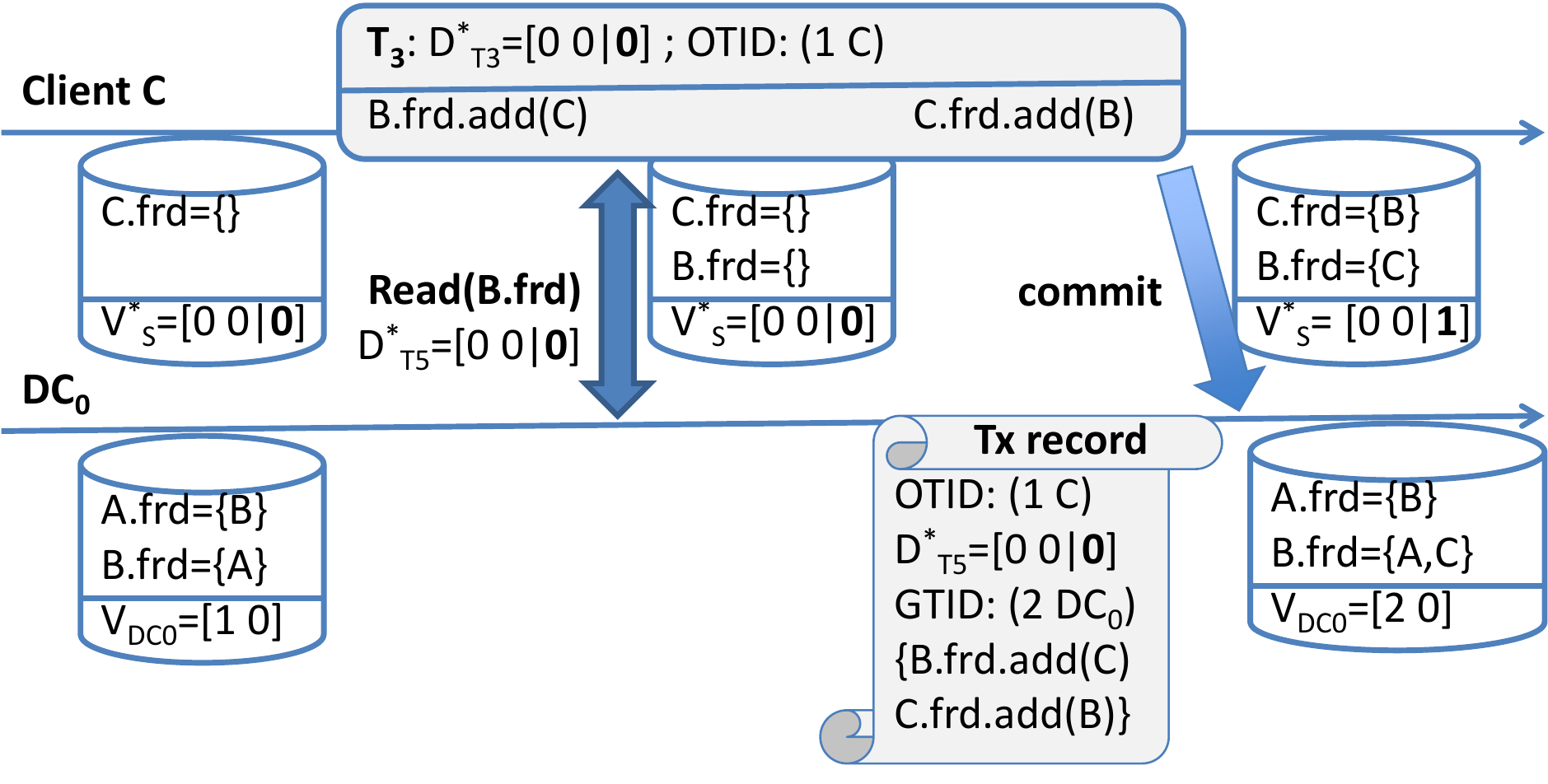}
  \squeezeAboveCaption
  \caption{Execution of $T_3$ from Figure~\ref{fig:transactions:newdeps}.
  \comment[Marek]{The graphical convention on this figure is quite unclear,
  and requires either improvement or explanation, or both.}}
  \squeezeBelowCaption
  \label{fig:transactions:run}
\end{figure}

\subsubsection{Transaction execution at scout}

An applications starts a transaction by executing 
\code{begin}.
This allocates a vector clock $\DStarTrans$ that summarises the causal
dependencies of the transaction.
It is set by default to the current state of the scout: $\DStarTrans :=
\VStarScout{S}$.
Concurrently, while the transaction executes, \VStarScout{S} increases as
$S$ processes committed transactions but \DStarTrans{} does not change.
\comment[Nuno]{If we want to get RC back:
Under Mergeable Snapshot Isolation, \DStarTrans{} does not change
thereafter.
Under Read Committed, it may increase to incorporate newly visible updates,
but never any faster than \VStarScout{S}.
}
Thus, $\DStarTrans \leq \VStarScout{S}$ is an invariant.

\code{begin} operation also generates the transaction's {Origin Transaction IDentifier}
(OTID), composed of a monotonically-increasing timestamp and the unique
scout identifier. 
Figure~\ref{fig:transactions:run} illustrates 
transaction execution with the run 
of transaction $T_3$ (from Figure~\ref{fig:transactions:newdeps}).
We will use this as a running example throughout this section.
In this example, OTID = \OTID{1}{$C$} and $\DStarTrans=[0,0|0]$ and
client C has the set of her friends, $\mathit{C.frd}$, on her cache.

Primitives \code{read} and \code{multi\_read} of the \SwiftCloud{} API,
read one or several objects respectively. 
They return a version of the requested object(s) that satisfies the
rules of \TransCausalPlus{}.
If the corresponding version is not found in the cache, it is fetched
from the DC\@.  \code{multi\_read} is an optimization that allows
to fetch multiple missing objects in a single round-trip.
If the corresponding version has been pruned away, the read fails, and the client has the
option of continuing or aborting the transaction with no effect.
\code{update} is called when an operation is executed on a previously
read object. 
In our running example, $T_3$ reads $\mathit{C.frd}$
from the cache but needs to fetch 
$\mathit{B.frd}$ from the DC before updating it. As $\DStarTrans=[0,0|0]$,
the version fetched from the DC still does not include the
updates from $T_1$, thus having $\mathit{B.frd}=\{\}$.

Reads and updates execute against the local copies returned by
\code{read{\slash}multi\_read}.
Update operations log their effect using the \code{update} primitive, so
that later, the updates can be transmitted to other replicas.

Our mergeable transactions are interactive, i.e., read sets and write sets are
determined on the fly by the transaction's control flow.
This enables, for instance, a user to display a consistent view of the
network of her friends.
This would be difficult in a system with non-interactive transactions
such as COPS \cite{rep:syn:1662} or Eiger \cite{syn:rep:1708}, since the
data to browse is not known until the transaction reads the user's
friend set.

When a mergeable transaction terminates successfully at the scout, it
commits locally.
If the transaction was read-only, the story stops there.
Update transaction logs its updates (if any) to durable storage, and
updates scout $S$'s own entry in \VStarScout{S} with the timestamp
part of the transaction's OTID\@.
The application may start a new transaction immediately.
Otherwise, its updates are now visible to later transactions at the
same scout.
\comment[NP]{it is necessary to fix this figure}
In the example of Figure~\ref{fig:transactions:run}, at the end of $T_3$,
we can see the cache with the updated value for $B.frd$ and $C.frd$ 
and the new value of
\VStarScout{S} = $[0,0|1]$, reflecting the OTID of $T_3$.

The scout globally-commits the mergeable transaction by sending
asynchronously the transaction's OTID, $\DStarTrans$ and updates to its
DC\@.
This asynchronous commit creates a durability issue, which we discuss in
Section~\ref{sec:FT}.
We expose the durability status of a transaction in the API so that the
application can make use of this information, as suggested by
\citet{rep:syn:1681-tr}.

\subsubsection{Transaction commit at the DC}

When a mergeable transaction is received in a DC, the DC first checks if
it satisfies the transaction's dependencies.
As long as the scout connects to the same DC, and the connection is FIFO,
this will be necessarily the case, since $\DTrans \leq \VScout{S} \leq
\VDC$, where \DTrans{} is the restriction of \DStarTrans{} to only DC
entries.
If not, the protocol waits until dependencies are satisfied.

Globally-committing a transaction by \DC{i} consists of the following
steps:
\begin{inparablank}
\item
  assign it a \emph{Global Transaction IDentifier (GTID)} \GTID{k}{i}
  such that $k=\Vi{i}[i]+1$,
\item
  log its commit record and update the DC replicas, and finally
\item
  increase $\Vi{i}[i]$ to $k$, thus making the transaction visible.
\end{inparablank}
The commit record contains the transaction's dependence vector
\DStarTrans{}, its OTID, its effects, and its GTID\@.
In our running example, $T_3$ is assigned GTID = \GTID{2}{0}, and
$\Vi{0} = [2,0]$ is updated accordingly.
The set of friends of B is updated by merging the new update into 
the DC version, by making $\mathit{B.frd} = \{A,C\}$.

Later, the DC sends the transaction commit record to other DCs using 
epidemic propagation \cite{rep:con:1414}.
A receiving DC logs the record durably, and if dependencies 
are satisfied, it applies the updates to the DC replicas
and updates its vector clock.
Otherwise, the DC delays processing the transaction until the 
dependencies are satisfied.

\subsubsection{Discussion}\label{sec:tx:discussion}

A globally-committed mergeable transaction (and the object versions that
it generates) is identifiable by both its OTID and GTID.
The OTID ensures uniqueness, and the GTID allows to refer to a
transaction efficiently in a dependence vector.
In some failure cases, a transaction may be assigned multiple GTIDs, but as
explained in the next section, they are treated equivalently.

Our protocol encodes the causal status of a whole node (DC or scout)
with a vector clock.
The scout-DC topology, the small and static number of DCs, and the
assumption that transaction processing in a DC is linearisable, 
all contribute to keeping these vectors small.
This very compact data structure summarises the whole past history,
i.e., the transitive closure of the current transaction's dependence.

%
%

\subsection{Cache maintenance}

A scout maintains a cache containing replicas of a subset of objects
(partial replication).
An application may ask to pin an object in the cache; otherwise, the
scout manages its cache with an LRU policy.
The cache may be updated, either as the result of a local transaction commit, or
because the cache is notified of a global commit by its DC\@.
A cache is always updated as the result of executing a stored transaction.

The partially replicated subset is guaranteed causally consistent, but not necessarily
complete.
For instance, assume the following update causal dependencies: $a
\leadsto b \leadsto c$; a scout might maintain only objects corresponding
to $a$ and $c$.
If a version (in this example, installed by update $b$) is missing in the cache,
it will be fetched from a DC, in accordance to the current transaction's snapshot.

The scout processes global-commit records in causal order and
atomically, ensuring that every state satisfies \TransCausalPlus{}.
The scout may either receive a full update on a cached object modified by the
transaction, in which case it installs a new version or just an invalidation.
A scout does not need to receive an update for an object that is not
cached; if it does, it is treated as a no-op.

\subsection{Implementation issues}\label{sec:tx:implementation}

\emph{Stored mergeable transactions:}
Stored mergeable transactions execute using the same approach as 
transactions executed in the scout, with the difference that
transactions access directly the replicas in the data centre
and that the clock of the data centre is used as the snapshot
vector.

\emph{Scaling transaction processing:}
The transaction processing process in the DC is linearisable as whole
and our DC implementation is parallel internally.
Data is partitioned across multiple storage nodes, and client
requests are processed by multiple proxy nodes.
Multiple global-commits can proceed concurrently.
It is only the update of the DC's vector clock that needs to be
linearised, since this is the step that renders a transaction visible.

A sequencer module at the DC sequentially assigns transaction
identifiers.
This could become a performance bottleneck as well as a central point of
failure.
Fault-tolerance can be improved by replicating the sequencer, for
instance by using chain replication \cite{pan:rep:1712} with a short
chain.
To improve performance, there could be more than one sequencer in each
DC, at the expense of larger vectors.

\emph{Security:}
\SwiftCloud{} caches objects and generates updates in the clients.
This poses no new major security threat, with access control enforcement at the
cloud boundaries addressing the problem. 

\SwiftCloud{} does not pose much of new challenges w.r.t.\ tolerating Byzantine
clients.
Incorrect operations can be tolerated similarly as in classic server-backed systems.
Forged dependence vector of a transaction cause no harm to other transactions,
same as client sending a wrong OTID (e.g., using the same OTID twice), as long as the DC
keeps track of GTIDs and a summary of updates related to the OTID to detect it.

\section{Fault-tolerant session and durability}
\label{sec:FT}

We discuss now how \SwiftCloudLong handles network, DC and client faults,
focusing on client-side mergeable transactions.
Our focus is primarily on our main contribution, mergeable transactions
executing on the client side; we mention other cases briefly.

At the heart of mergeable transactions is causal consistency (i.e.,
session guarantees \cite{rep:syn:1481,Brzezinski2004}), which is easily
ensured in the failure-free case, since DCs exchange transactions using causal
broadcast, a DC commits a single transaction at a time, and a scout connects to
a single DC over a FIFO channel.
However, when a scout loses communication with its current DC,
due to network or DC failure,
the scout may need to switch over to a different DC\@.
The latter's state is likely to be different, and it might have
not processed some transactions observed or indirectly observed (via
transitive causality) by the scout.
In this case, ensuring that the clients' execution satisfies the consistency model
and the system remains live is more complex.
As we will see, this also creates problems with durability and
exactly-once execution.

As a side-effect of tolerating DC faults and fail-over, our protocols
also support client disconnection.
Obviously, if a disconnected client requires state that is not currently
cached, it cannot make progress.
Similarly, if the client remains permanently disconnected or loses its
durable state before reconnecting, there is not much that can be done.
However, in-cache disconnected operation is supported ``for free,'' as
long as the scout remains live and eventually reconnects.

\subsection{Durability and exactly-once execution issue}

The scout sends each transaction to its DC to be globally-committed, to
ensure that the DC stores it durably, allocates a GTID, and eventually
transmits it to every replica.
If it does not receive an acknowledgment, it must retry the
global-commit, either with the same or with a different DC\@.
However, the outcome of the initial global-commit remains unknown.
If it happens that the global commit succeeded with the first DC, and
the second DC allocates a second GTID, the danger is that the
transaction's effects could be applied twice under the two identities.

For some data types, this is not a problem, because their updates are
idempotent, for instance \code{put(key,value)} in a last-writer-wins
map.
For other mergeable data types, however, this is not true: think of
executing \code{increment(10)} on a counter.
Systems restricted to idempotent updates can be much simpler
\cite{syn:rep:1708}, but in order to support general mergeable objects
with rich merge semantics, \SwiftCloud{} must ensure exactly-once
execution.

In principle, the DC could check that the transaction's unique OTID does
not appear in the log.
Unfortunately, this is insufficient, since the log might  be
pruned while a scout was disconnected (e.g., during a long journey).

\subsection{Causal dependency issue}

When a scout switches to a different DC, the state of the new DC may be
unsafe, because some of the scout's causal dependencies are missing.
For instance, in Figure~\ref{fig:architecture}, suppose that transaction
\GTID{23}{2} created some object, and later, the Paris scout updates
that object.
Vector clock $\Vi{O}[2]=22$ reveals that \DC{0} has not yet processed the
creation transaction.
Therefore, committing the update transaction to \DC{0} would lead DC
into an unsafe state.
Unless the scout can find another DC that has processed all the
transactions that it depends upon, its only option is to wait until
\DC{0} receives \GTID{23}{2}.
This might take a long time, for instance if \DC{2} (which committed the
missing transaction) is unavailable.
\comment[MZ]{This example is not very food, as updated object is easily
recoverable by client.}

Before presenting our solution in the next section, let us consider some
possible approaches.

Some geo-replication systems avoid creating dangling causal dependencies
by making synchronous writes to multiple data centres, at the cost of
high update latency \cite{rep:pan:1693}.
Others remain asynchronous or rely on a single DC, but after failover
clients are either blocked like in our example (unavailability) or they 
violate causal consistency \cite{rep:syn:1662,syn:rep:1708,rep:syn:1690}.
The former systems trade consistency for latency, the latter trade latency for
consistency or availability.

An alternative approach would be to store the dependencies on the scout.
However, since causal dependencies are transitive, this might include a large
part of the causal history and a substantial part of the database.\footnote{
Requiring programmer to provide explicit causal dependencies at the programming
level may reduce the amount of direct dependencies \cite{explicit-socc2012},
nevertheless indirect dependencies are still of a problem.
}
It would be similar to a peer-to-peer system, every scout being a
first-class full replica, with its own component in vector clocks.
Traffic and storage requirements would be unbearable in this case.

Finally, a trivial solution would be for a client to observe only its own
updates.
This would ensure safety but, lacking liveness, would not be useful.
To exclude such trivial implementations, we impose the convergence
requirement that a client eventually observes all committed updates.
Such a relatively weak property does not preclude serving the client with
an old safe version, the freedom we use in our approach.


\subsection{Fault-tolerant causal consistency}
\label{sec:FT:causal}

Our approach is to make scouts co-responsible for the recovery of
missing session causal dependencies at the new DC\@.
Since, as explained earlier, a scout cannot keep track of all
transitive dependencies, we restrict the set of dependencies.
We define a transaction to be \emph{$K$-durable} at a DC, if it is known to
be durable in at least $K$ DCs, where $K$ is a configurable
threshold.
Our protocols let a scout observe only the union of:
\begin{inparaenum}[\it (i)]
\item 
  its own updates, in order to ensure the ``read-your-writes'' session guarantee
  \cite{rep:syn:1481}, and
\item
  the {$K$-durable updates} made by other scouts, to ensure other session
  guarantees, hence causal consistency.
\end{inparaenum}
In other words, the client depends only on updates that the scout
itself can send to the new DC, or on ones that are likely to be found in
a new DC\@.
The set of $K$-durable updates is causally consistent, i.e., it is
transitively closed over the $\leadsto$ relation.
When failing over to a new DC, the scout helps out by checking whether the
new DC has received its recent updates, and if not, by repeating the
commit protocol with the new DC\@.

The scout can switch to a DC, as long as this new DC ensures that the
scout continues to observe a monotonically-growing set of $K$-durable
updates.
This is possible, since the scout's own updates that are not $K$-durable
cannot depend on updates from another scout that are themselves not
$K$-durable.

\SwiftCloud{} prefers to serve a slightly old but $K$-durable version,
instead of a more recent but more risky version.
Instead of the consistency and availability vs.~latency trade-off of
previous systems, \SwiftCloud{} trades availability for staleness.
Since our system relies on gossiping between DCs, to some extent, the
larger the parameter $K$, the higher the probability that an update that
is $K$-durable at some DC will be found $K$-durable in another random DC\@.
However, higher values of $K$ cause updates to take longer to become
visible and may increase the risk that an update is blocked by a network partition.
On the other hand, lower values may increase the chance that a scout
will not be able to find a suitable DC to fail-over to.
In particular, $K = 1$ corresponds to the original blocking session-guarantees
protocol of \citet{rep:syn:1481}.

In practice, $K \ge 2$ is a good compromise, as it ensures session
guarantees without affecting liveness in the common case of individual
DC failures or disconnections \cite{rep:pan:1693}.
Our implementation uses $K = 2$, tolerating a single fault.
A better approach might be $K=3$, to tolerate a single fault when a DC
is closed for scheduled maintenance \cite{rep:pan:1693}.


By delaying visibility, rather than delaying writes like some previous works,
we move the cost of causal consistency from the domain of commit-time latency,
into the domain of data staleness.
Our evaluation in Section~\ref{sec:eval} shows that our approach 
improves latency, with a negligible impact on staleness.
The staleness increases concurrency of updates in the system, which is
tolerable, since \SwiftCloud uses mergeable objects to handle that seamlessly.

\subsubsection{Discussion}

\citet{syn:rep:1677} establish that causal consistency%
\footnote{
  Specifically, a stronger variant involving real-time dependencies.
}
is the strongest achievable consistency in an always-available
convergent system.
The practical problem of ensuring similar guarantees in the presence of
partial replicas, or of clients switching servers, was not addressed before.
We demonstrated how to ensure \TransCausalPlus{} for clients under this
new assumption, at the price of weaker liveness property.
  
\subsection{Fault-tolerant exactly-once execution}
\label{sec:ft-exactly-once-solution}

We now address the remaining issue of ensuring that each update is delivered
exactly once at each replica, a problem that arises with any commit protocol
that allows retries.
Simply repeating the global-commit protocol until the scout receives an
acknowledgment takes care of one half of the problem, i.e.,
at-least-once delivery.
We now consider the other half, eliminating duplicates in the presence
of failures and pruning.

Our approach separates the concerns of tracking causality and of uniqueness,
following by the insight of \citet{dotted-vv}.
Recall (Section~\ref{sec:tx:discussion}) that a transaction has both a
client-assigned OTID, and one or more DC-assigned GTIDs.
The OTID identifies it uniquely, whereas a GTID is used when a summary of
a set of transactions is needed.
Whenever a scout globally-commits a transaction at a DC, and the DC does
not have a record of this transaction already having a GTID, the DC
assigns it a new GTID\@.
This approach makes the system available, but may assign several GTID aliases
for the same transaction.
All alias GTIDs are equivalent in the sense that, if updates of $T'$
depend on $T$, then $T'$ comes after $T$ in the causality order,
no matter what GTID $T'$ uses to refer to $T$.

When a DC processes a commit record for an already-known transaction with
a different GTID, it adds the alias GTID to its commit record on durable
storage.

To provide a reliable test whether a transaction is already known,
each DC maintains durably a map of the last OTID received from each scout,
noted \maxOTIDs{i}{S} for scout $S$ and \DC{i}.%
\footnote{
  The number of entries in \maxOTID{}{} is the number of
  scouts, which can be large.
  However, a map is local to a DC, and never transmitted and the number of
  active clients is more limited.
  Supporting even millions of clients is well within the DC storage
  capabilities.
}
Thanks to causal consistency, \maxOTIDs{i}{S} is monotonically non-decreasing.

When a \DC{i} receives a global-commit message from scout $S$, it checks
that its OTID is greater than $\maxOTIDs{i}{S}$; if so, it allocates a new
GTID, logs the commit record, and returns the GTID to the scout.
If it is not greater, this means that this transaction has already been
delivered to this DC\@.
In this case, the DC searches its log for a commit record with the same
OTID.
If one is found, the DC returns the corresponding GTID to the scout.

Otherwise, this means that the commit record has been pruned.
This raises the question of how the client will refer to the transaction
in causal dependencies of subsequent transactions.
It turns out this is not necessary: as only transactions that were
processed by all DCs are pruned, such dependencies will be always
satisfied; therefore, a null GTID is returned to the client.

Note that scouts do not need to worry about exactly-once delivery, since
a scout will communicate with a DC only if the latter has processed a
superset of the $K$-durable transactions that the scout has observed.

\subsection{Fault tolerance on server-side}

The fault tolerance algorithm just described is not directly applicable to update
transactions executing on the server (DC) side, issued by
\emph{exec\_stored\_proc} call.
 We discuss briefly possible fault tolerance options here; these were not
 implemented in our prototype.

In the case of non-mergeable transactions, it is sufficient to tag a request with
OTID and eliminate duplicate execution using existing concurrency-control,
treating duplicates as concurrent conflicting transactions.

A similar technique can be applied to mergeable transactions executing
on the server-side.
In this case, however, the OTID is augumented with a dependency vector $\DStarTrans$.
As long as client uses the same OTID and $\DStarTrans$ for reissued
transaction requests and the transaction processing is deterministic
w.r.t.\ a database version, duplicate execution of the transaction can only
result in producing updates with the same identity, which is addressed
by the technique in Section~\ref{sec:ft-exactly-once-solution}. 

\cut{Mergeable update transactions executing on the server-side need to be treated
differently than their client-side counterparts.
Similarly to client-side execution, client can fail to receive an outcome of a
mergeable transaction request and repeat the transaction request without knowing
the original request status.
A transaction request may be non-idempotent too, in which case
each server-side transaction execution generates different database updates,
leading to a potential consistency issue.
To avoid the problem, we can let client assign transaction OTID and a dependencies
vector $\DStarTrans$ at the time of remote update transaction request generation.
As long as client uses the same OTID and $\DStarTrans$ for reissued
transaction requests and the transaction processing is deterministic
w.r.t.\ to a database version, duplicate execution of the transaction will
result in producing updates with the same identity.
Duplicated updates with the same OTID are harmless thanks to the approach introduced in
Section~\ref{sec:ft-exactly-once-solution}.
}


\section{Building applications}    
\label{sec:apps}

The \SwiftCloud approach is designed to fit best applications with sufficient
locality to run in a small cache even if the total database size is large, and
that use mergeable transactions mostly.
We demonstrate the applicability of our application and consistency model by
implementing a range of applications that meet these characteristics.
Evaluating the performance of these applications is the focus of
Section~\ref{sec:eval}.

\subsection{\SwiftSocial social network}
\label{sec:apps:social}

The \SwiftSocial application is our port of WaltSocial, a simple social
network prototype \cite{rep:syn:1661}.
The port was straightforward, using the data types from the CRDT library and
transactions.
The \SwiftSocial-specific code consists of approximately 680 Java LOC with
few comments.

\SwiftSocial maps (using a CRDT map) a user to his profile information (a
LWW-Register) and to set CRDTs containing his wall messages, events, and
friendship requests.
The event set records every action involving the user, and thus
grows linearly in the number of updates.

Update transaction types include registering a user, login (fetches the
user's profile and checks his password; subscribes to updates), logout
(unsubscribes), posting a status update on the user's wall, and sending
a message to another user's wall or accepting friendship request.
Read-only transactions view another user's wall, or list his friends.
The workloads used in Section~\ref{sec:eval} consist of user sessions
running a mix of these transactions.

Transaction's atomicity ensures that accepting a friendship request updates
both users' friendship sets consistently or a notification event is added
together with a wall post.
Causality naturally helps user-experience, for example when a user follows
a conversation thread, replies appear after the original messages.
It also helps programming in some cases, since, in the same case, reply to
a message does need to be processed (e.g. rendered) without a message.

We found that only the user registration transaction should preferably be of a
non-mergeable type, in order to name users uniquely.
A more advanced social network application could benefit from some server-side
procedures, e.g., for searching user, content or suggesting advertisements.  


\subsection{\SwiftFS collaborative documents}
\label{sec:apps:tree}

Collaboration tools are typical applications that benefit from
low-latency, highly-available client-side access to data, including
offline.
Our \SwiftFS{} application implements a mergeable file-system hierarchy
with mergeable file types that allows to share documents and edit them
concurrently.
\SwiftFS subsumes some of the functionality of a DVCS like Git or
Mercurial, and of online collaborative editors such as Google Docs.

\SwiftFS consists of a naming tree of \emph{directories},
implemented using CRDT map objects.
A directory maps unique keys, which are strings of the form
\filename{name}{type}, to objects of arbitrary CRDT type.
This hierarchical structure provides a variety of semantically different
data objects for collaboration.
For instance, \filename{foo.txt}{lww} refers to a file managed as
untyped blob with LWW-Register semantics, whereas
\filename{foo.txt}{seq} is a text file with fine-grained automatic
merging of updates, managed as a conflict-free sequence
\cite{app:rep:1652}.
A family can edit their family tree without conflict by storing it as a
CRDT object of type graph, e.g.,
\filename{The\_Simpsons\_family\_tree}{graph}.

Transactions ensure consistency across multiple object and updates.
For instance, a user might snapshot a subtree, perform updates
throughout it (e.g., replace the word ``\SwiftCloud{}'' in all files
with another name), copy it to a different place, and delete
the original, all as a single transaction. 



Concurrent updates are merged using the following heuristic.
Concurrently creating and populating two directories under the same name
takes the union of elements; elements with the same \filename{name}{type}
are merged recursively.
The semantics of merging two files is given by the \code{merge} method of
their type.
Embedding the type in the lookup key ensures that only files  of the
same type are merged.

Removing a directory recursively calls the \code{remove} interface of its
elements.
If \code{remove} is concurrent with another user's update, remove wins and
update is lost.
However, the second user's work can be accessed by recreating a snapshot that
does not include the deletion.
Furthermore, she can reinstate the missing file under its original
unique key using a version of \code{create()}.

The \code{move} operation is implemented as a mergeable transaction that
takes a snapshot of the source subtree, copies it recursively, then
recursively removes the source subtree.
Anomalies may occur if the source is modified concurrently.
To implement the Posix \code{rename} semantics, re-linking a
subtree at a different location, would require using a
classical serialisable transaction; otherwise cycles might
appear~\cite{syn:rep:1652}.

\SwiftFS operations exhibit good locality: caching a  subtree and a global
map may be often sufficient for good latency and disconnected support.

\subsection{TPC-W benchmark}
\label{sec:apps:TPCW}

The TPC-W benchmark simulates an online book store.
Simulated clients browse items, interact with shopping carts, and check
out.
We ported an existing open-source Java implementation \cite{tpcw-impl}
to \SwiftCloud.
Our objectives were to demonstrate porting an existing application, and
to provide the reader of this paper with a familiar point of reference.

Transactions are essential in TPC-W but most can be mergeable, with
checkout being an exception that needs synchronous execution in the DCs. 
Checkout atomically pays for, and adjusts the stock of, each item 
in the shopping cart.

We model TPC-W database with CRDTs.
For instance, a CRDT set represents the shopping cart, thus avoiding the
anomalies of the Amazon shopping cart \cite{app:rep:optim:1606}.
We index product records using CRDT sets.
A CRDT counter is used to track the stock of each item.
The benchmark specification allows stock to become negative, as long as
is eventually replenished.
If desired, enforcing non-negative stock could be achieved by using
non-mergeable transactions.

TPC-W can display some problems when naively ported to being executed at the edge.
Namely, operations involving large read sets, such as queries to the whole product database, 
tend to perform poorly in the client cache. \SwiftCloud can tackle this sort of issue by leveraging stored transactions.


\section{Evaluation}    
\label{sec:eval}


This section presents an experimental evaluation of \SwiftCloud based on
the applications described in the previous section. The aim of this study is to assess the relative strengths and weaknesses of
executing application logic at the two opposite ends provided by the \SwiftCloud platform. Namely,
we will compare our caching approach, executing both
reads and updates at the client, against the standard approach of doing
updates in the DC\@. As such, the horizontal scalability of DCs is not evaluated; in fact, all
the (parallel) components of a \SwiftCloud{} DC run on a single server
in these experiments.

\begin{figure}[tp]
  \centering
  \includegraphics[width=0.7\columnwidth]{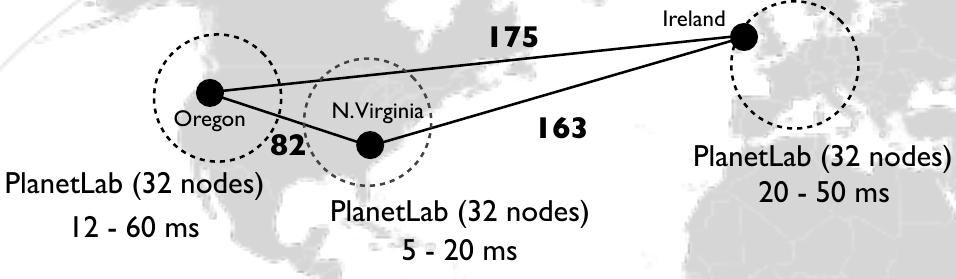}
  \squeezeAboveCaption
  \caption{Experimental topology and round-trip times.}
  \squeezeBelowCaption
  \label{fig:topology}
\end{figure}

\subsection{Experimental setup}

\begin{figure*}[t]
\centering
\subfigure[All clients]{
  \includegraphics[width=0.48\textwidth]{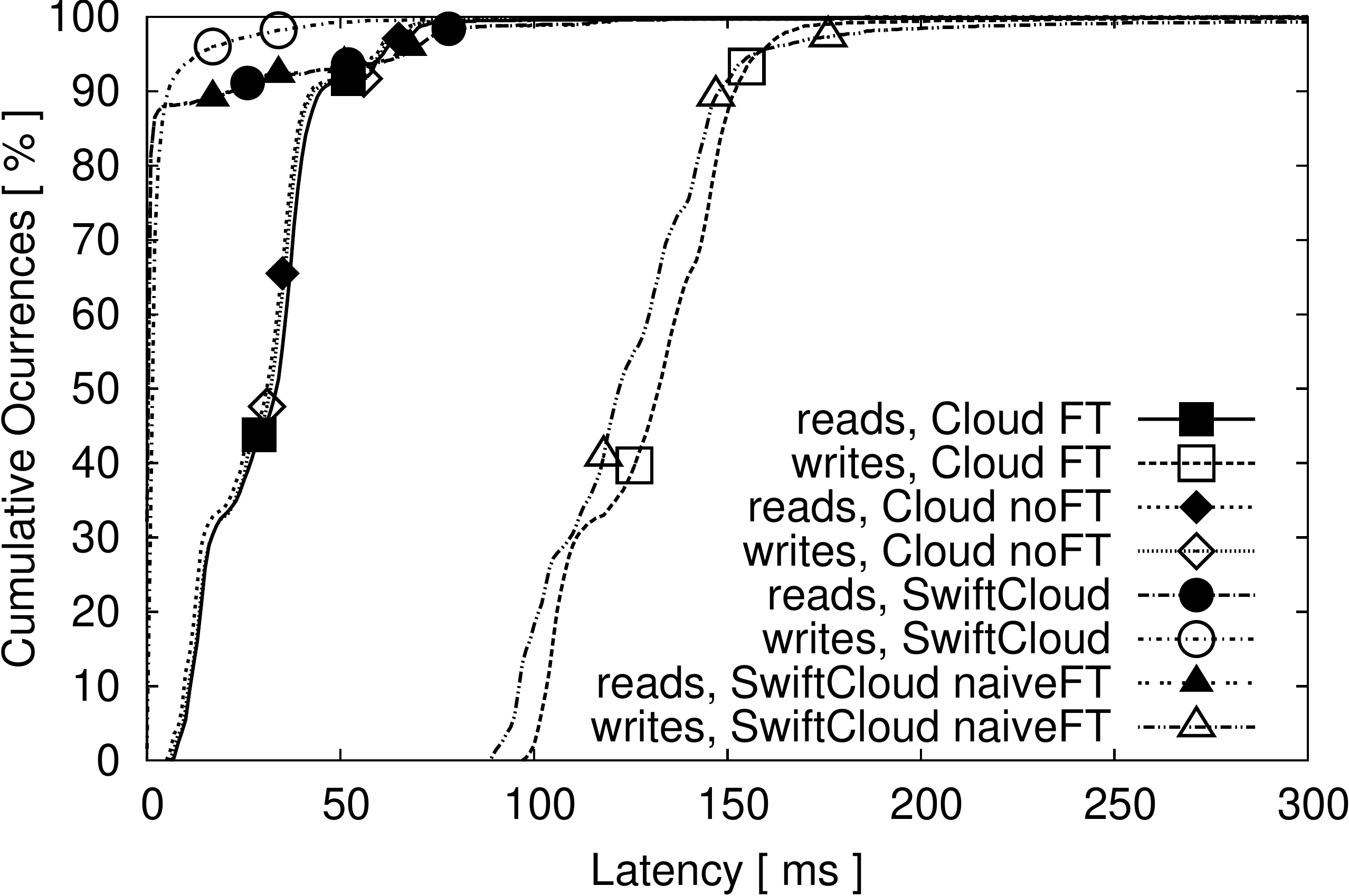}
  \label{fig:latency:global}
\squeezeBelowCaption}
\hspace{-3mm}
\subfigure[One client]{
  \includegraphics[width=0.48\textwidth]{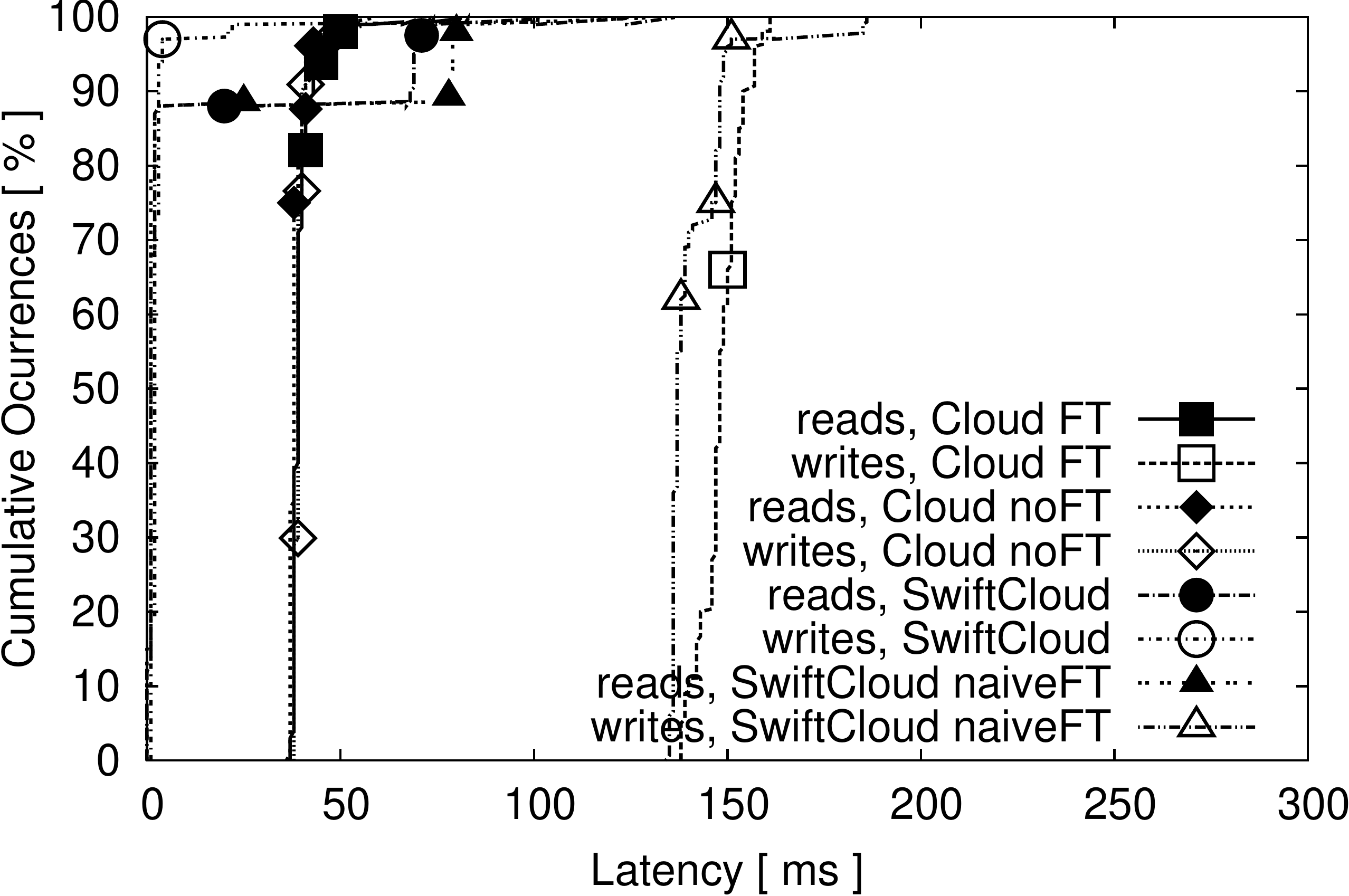}
  \label{fig:latency:one}
\squeezeBelowCaption}
\hspace{-3mm}
\subfigure[Various cache miss-ratios]{
  \includegraphics[width=0.48\textwidth]{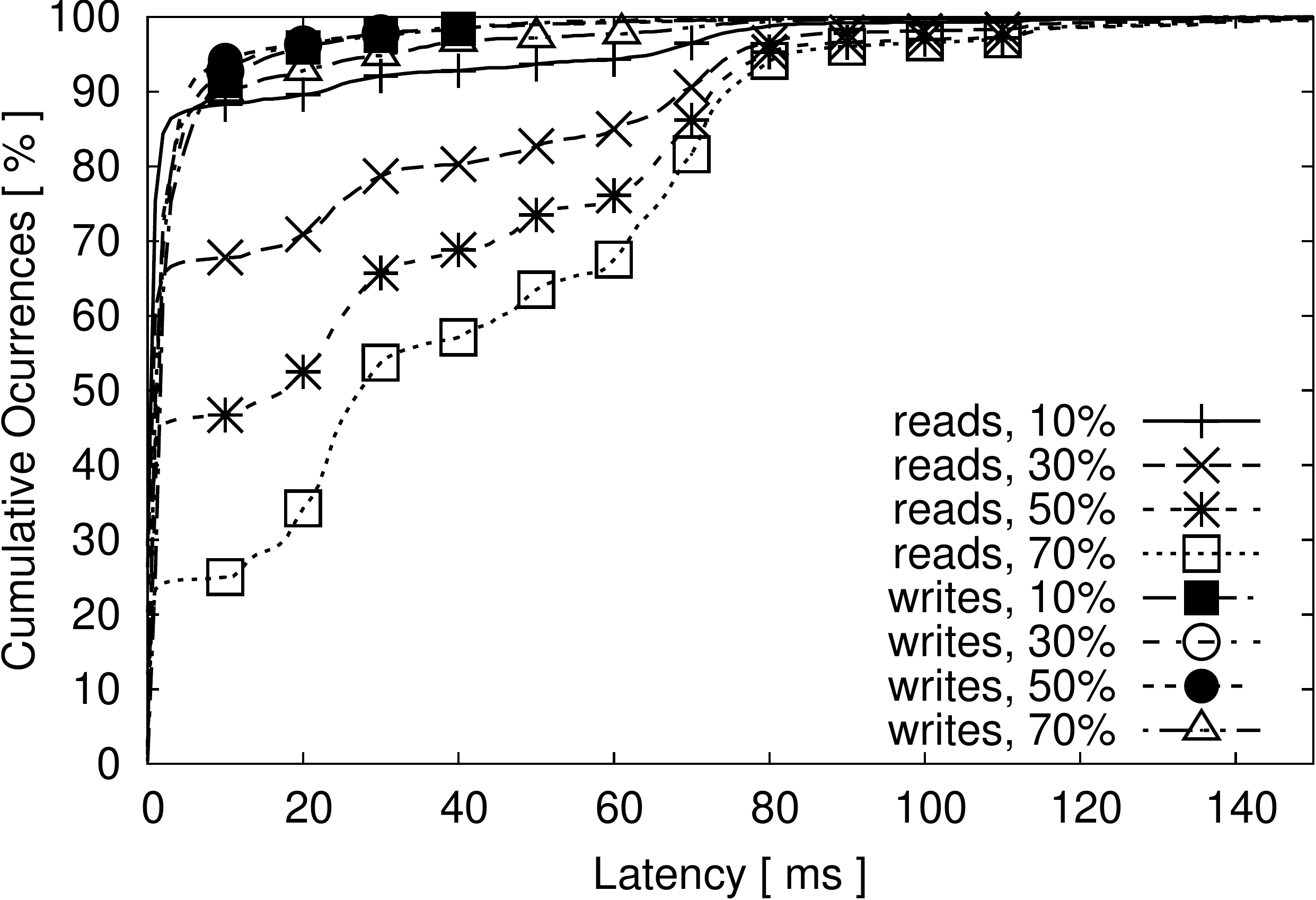}
  \label{fig:latency:workload}
\squeezeBelowCaption}
  \squeezeAboveCaption
\caption{Perceived latency for \SwiftSocial.}
  \squeezeBelowCaption
\end{figure*}

\SwiftCloud is written in Java.
Approximate code sizes, including javadoc, are as follows:  
whole system, 20K LOC; DC-specific, 3.5K LOC; 
scout-specific, 3K LOC; CRDT library 5K LOC. 
It runs over a pre-existing communication, serialisation, 
and DHT package of approximately 12K LOC. 
Durable state is stored in a Berkeley DB database.

We run DCs in three Amazon EC2 availability zones, and clients on
96~PlanetLab machines located geographically near the DCs.
Figure~\ref{fig:topology} describes their approximate geographical
locations and round-trip times (RTTs).
EC2 machines are equivalent to a single core 64-bit 2.0 GHz Intel Xeon
virtual processor (2\,ECUs) with 3.75 GB of RAM and run Java OpenJDK
64-bit IcedTea7 (Server) above Linux 3.2.
PlanetLab nodes have heterogeneous specifications and latencies.
We use default system settings throughout.

We compare configurations with {one DC} (Ireland), two (+\,Oregon)
and three DCs (+\,North Virginia). 
Within an experiment, we vary parameters but keep constant the set of
PlanetLab nodes.
We vary the number of clients by adding more independent client threads per
PlanetLab node, thus keeping the network latency distribution invariant.

A state-of-the-art  geo-replication configuration is achieved by
co-locating \SwiftCloud scouts (with a cache size set to zero) within the DC\@.
Its non-fault-tolerant ``\SOA-noFT'' configuration performs its updates
at a single DC synchronously, and propagates them asynchronously to the
others.
The ``\SOA-FT'' configuration writes to two DCs synchronously, simulating
disaster-tolerant geo-replication systems such as a configuration of
Walter \cite{rep:syn:1661}.   

In the fault-tolerant ``\SwiftCloud{}'' configurations, 
each client thread has a
co-located scout, with a dedicated cache size of 512~objects, and uses
the asynchronous global-commit protocol.
In the alternative {``\SwiftCloud-naiveFT''} configuration, commit is synchronous
and returns only when durable in at least two DCs.

All configurations evaluated, including \SOA-FT and \SOA-noFT, leverage the same codebase. They represent
the extremes of the \SwiftCloud system application logic distribution spectrum. The \SOA-{FT/noFT} configurations correspond to the classical
approach where most or the entire application logic runs at the server in the DC, whereas the \SwiftCloud configurations 
strive for the opposite, moving as much as possible to the client. Our evaluation emphasises the impact on 
performance of these two opposites.

\subsection{Latency}\label{sec:eval:latency}

We first evaluate the responsiveness of end-user operations, using the
\SwiftSocial benchmark.
It simulates 25,000 users, each one associated to 25 friends uniformly
at random, simulating user sessions as described in
Section~\ref{sec:apps:social}.
10\% of transactions involve modifications; the rest are read-only.
90\% of transactions involve data of the current user and his
friends, and hit the cache once it is warm; the other  10\% target
 (uniformly) random users and produce cache misses.


Figure~\ref{fig:latency:global} plots the CDF of the perceived latency of
executing a transaction.
In the \SwiftCloud default configuration, around $90\%$ of transactions have
near-zero latency, the remaining $10\%$ having variable latencies.
This corresponds nicely to the 90{\slash}10\% that respectively
hit{\slash}miss in the cache; the cost of a miss depends on the RTT to
the closest DC, which varies between PlanetLab nodes.
Remember that, with mergeable data, updates occur in cache, not just
reads.

In the \SOA-noFT configuration, transaction latency is proportional to
the client-DC RTT\@.
\SOA-FT suffers additional latency for writing to a quorum.
These classical configurations provide worse latency for both reads
and writes, when compared with \SwiftCloud.
Fault-tolerant approaches requiring writing to a quorum of replicas
synchronously penalise writes heavily,
when compared with \SwiftCloud client assisted failover approach.
The same happens with the \SwiftCloud{-naiveFT} 
configuration for the same reason.


Figure~\ref{fig:latency:one} shows the operation latency experienced by
a particular client (other clients have a similar pattern, with the
lines being shifted right or left depending on their RTT to the DC).

In \SOA-noFT and \SOA-FT, submitting a request costs 
a single RTT to the DC.
With the \SwiftCloud approach, each miss costs one RTT\@.
When the benchmark accesses a non-friend, a cache miss fetches the first
read, followed by several others (usually a \code{read} followed by a
\code{multi\_read}). For applications relying highly on client-side
execution, the main drawback is that cache misses can be costly. It can be mitigated by
moving execution of the most offending code paths to the server.   
To further address this issue in systematic a manner, we are implementing a 
mechanism to automatically switch to DC-side execution upon a cache miss.

\begin{figure}[t]
  \centering
  \includegraphics[width=0.48\textwidth, height=5cm]{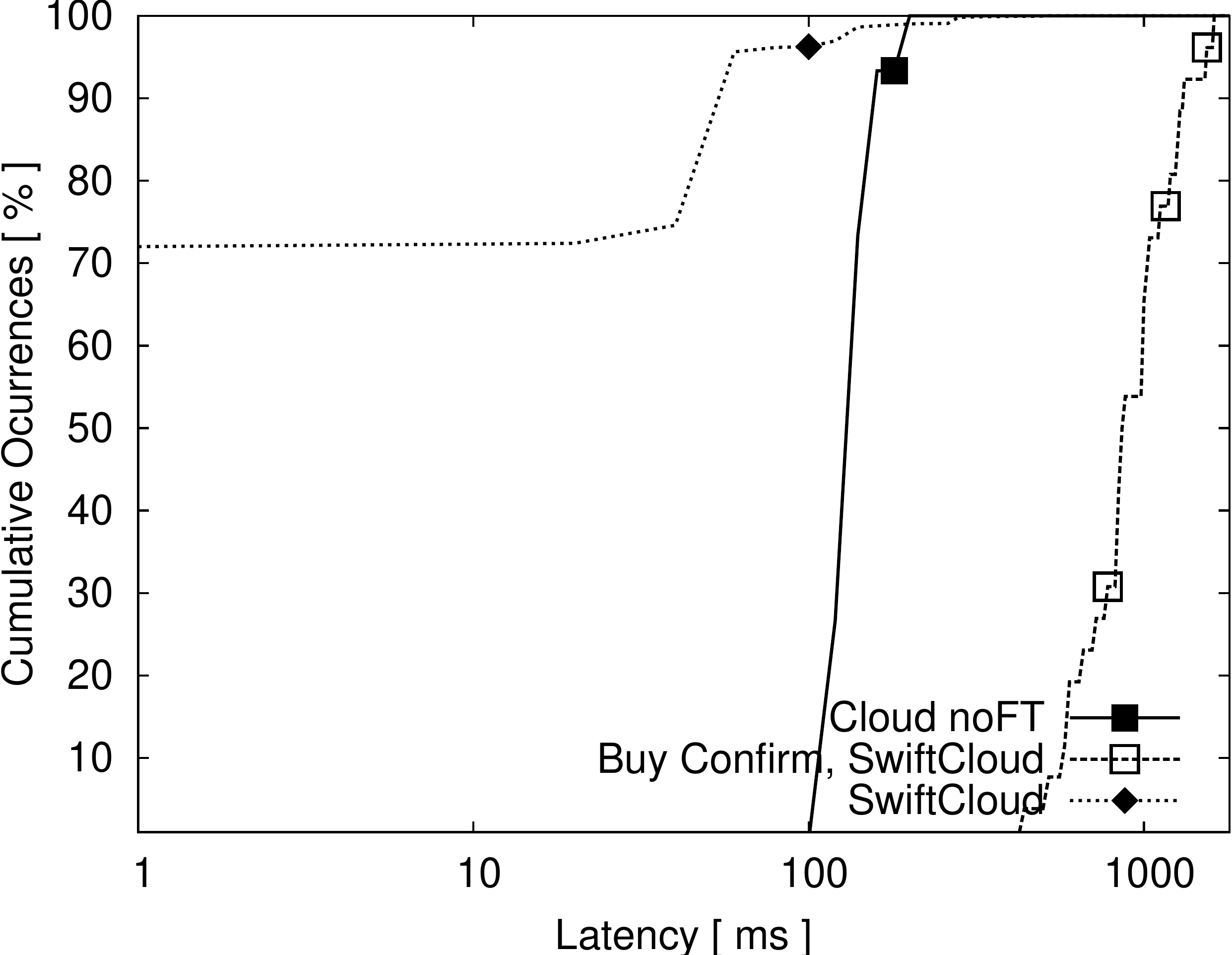}
  \label{fig:latency:notify}
\squeezeAboveCaption
  \caption{Latency for TPC-W browsing workload.}
\squeezeBelowCaption
  \label{fig:latency:tpcw}
\end{figure}

Figure~\ref{fig:latency:workload} plots the CDF of the perceived latency increasing
the cache miss ratio by increasing the ratio of operations over non-friends. 
We can see that $90\%$  of read-write transactions have zero latency, as
writes are always on cached data, i.e., objects from the user or from friends.
The ratio of read-only transactions experiencing zero latency is directly 
proportional to the cache hit ratio, as cache misses must be served from the DC. 
These results are compatible with the results obtained for TPC-W
browsing workload with $95\%$ read-only transactions, in a 
similar deployment, and a system configuration that exhibits 
$76\%$ hit ratio (Figure~\ref{fig:latency:tpcw}).
This result also shows that the checkout operation (\emph{buy confirm}),
requiring synchronous execution in the DC, presents high latency, as expected.

\begin{figure*}[t]
\centering
\subfigure[$10\%$ cache misses]{
  \includegraphics[width=0.48\textwidth]{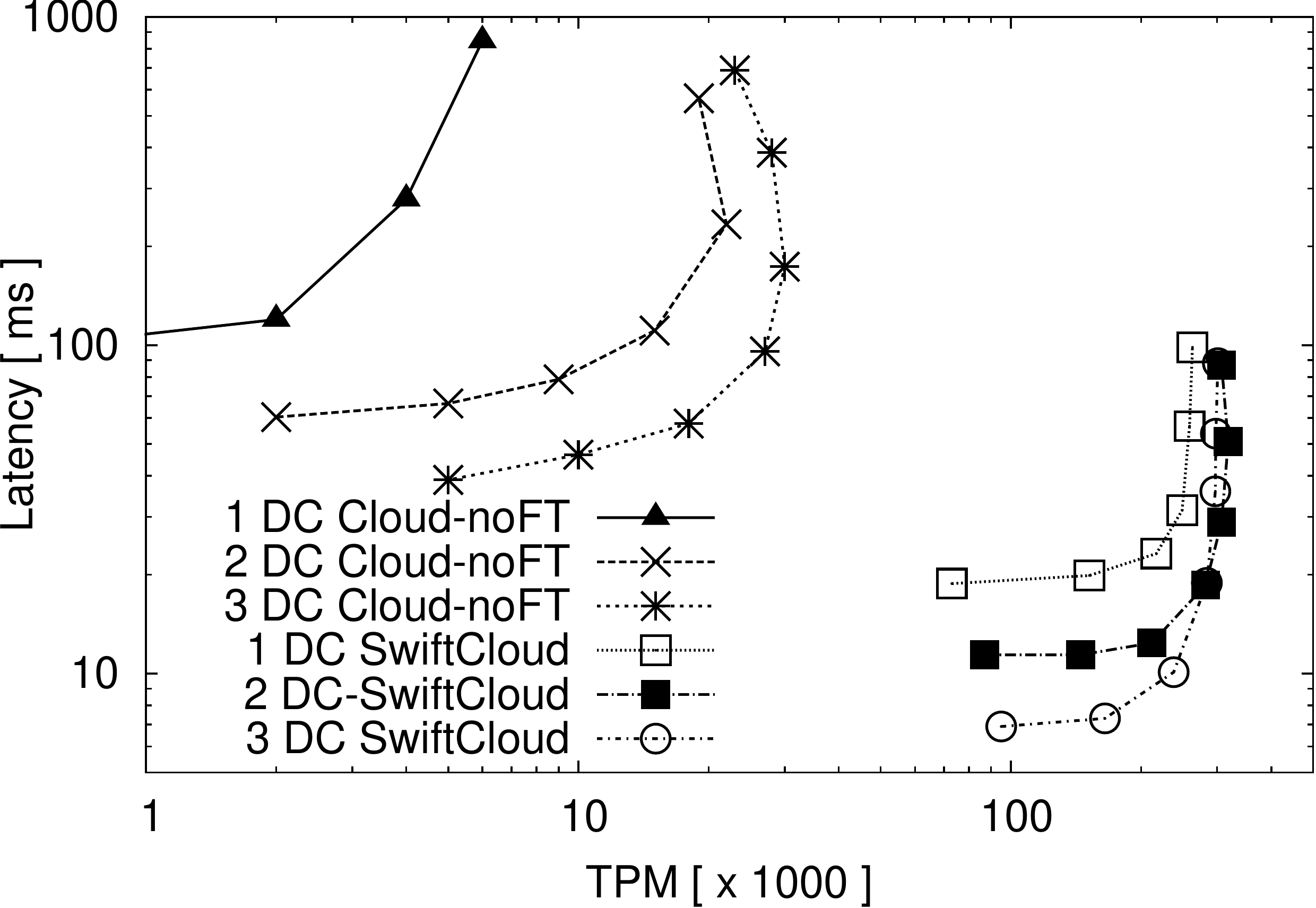}
  \label{fig:swiftsocial:perf:9010}
\squeezeBelowCaption}
\subfigure[$50\%$ cache misses]{
  \includegraphics[width=0.48\textwidth]{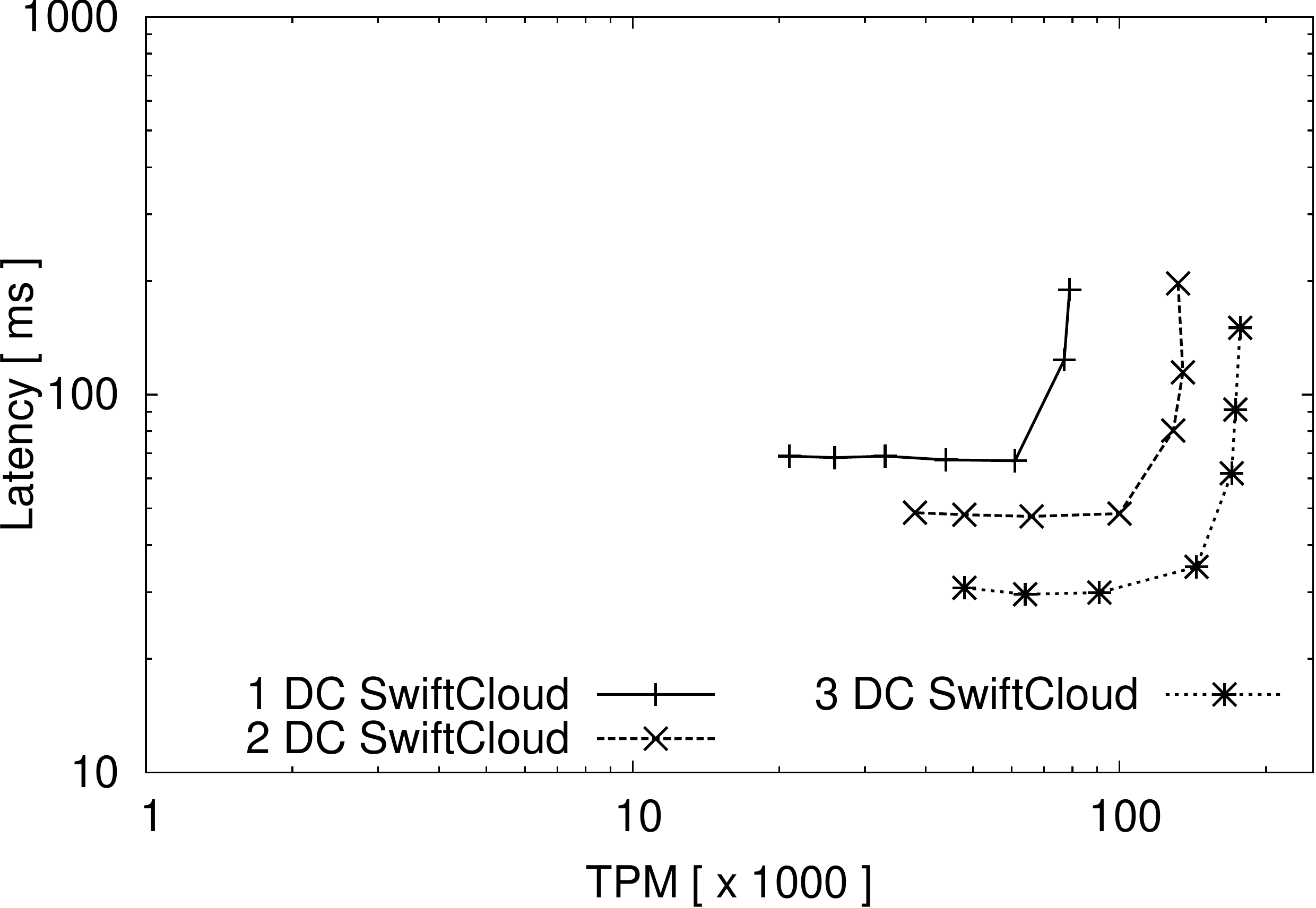}
  \label{fig:swiftsocial:perf:5050}
\squeezeBelowCaption}
\squeezeAboveCaption
\caption{Performance of \SwiftCloud (\SwiftSocial).}
\squeezeBelowCaption
\end{figure*}

\subsection{Throughput vs.~latency}

We now investigate how \SwiftCloud performance compares with classical
geo-replication approaches and how it scales with the number of DCs.
Figure~\ref{fig:swiftsocial:perf:9010} plots throughput vs.~latency,
comparing \SwiftCloud{} with \SOA-noFT, running the
same \SwiftSocial{} benchmark as before.
It shows configurations with one, two and three DCs.
In each configuration, we increase the load by adding more simulated
clients.
As the load increases, initially throughput improves while latency does
not change; however, as the system saturates, throughput ceases to
improve and latency becomes worse.
Eventually, throughput decreases as well.
We use a log-log scale; down and to the right is better.

The plot shows that, at equal hardware cost, \SwiftCloud has
order-of-magnitude better response time and better throughput, than the
classical geo-replication approach, even though \SwiftCloud{} is fault
tolerant and \SOA-noFT is not.
The explanation is simple: \SwiftCloud{} absorbs 90\% of transactions in
its cache.
Recall that even updates are cached.

Interestingly, although adding a third DC to \SwiftCloud{} improves
latency and throughput at first, it does not improve peak performance at
saturation, in contrast to the DC-based approach.
The reason is that DCs are fully replicated, i.e., every DC processes
every update transaction.
In the \SOA case, additional DCs allow to process more read-only
transactions in parallel, but for \SwiftCloud{} this effect is
negligible because read-only transactions were already absorbed by
the client-side cache.
In this benchmark, 10\% of the transactions are read-only transactions
that access the DC because of a miss, and 10\% are updates that must
global-commit in the DC\@.
With a single DC, there is an equal number of both.
With two DCs, each DC processes only half of the reads but all of the
updates.
Thus, the impact of adding a DC is less than in the \SOA setup.
Additionally, the faster read transactions execute, the faster
additional commits are sent to the DCs.
This trend continues when adding additional DCs.

We confirm this explanation by increasing the amount of cache misses to
$50\%$ in Figure~\ref{fig:swiftsocial:perf:5050}.
The ratio is now five read-only transactions for one update transaction.
This larger ratio is expected to enable \SwiftCloud to scale with the
number of DCs, as more read-only transactions will benefit from
executing in a closer DC.
This hypothesis is confirmed by the plot.

In summary, client-side caching of mergeable data enables scalable
shared storage with a potentially reduced, cheaper DC infrastructure.


\begin{figure*}[t]
\centering
\includegraphics[width=0.58\textwidth,height=4.0cm]{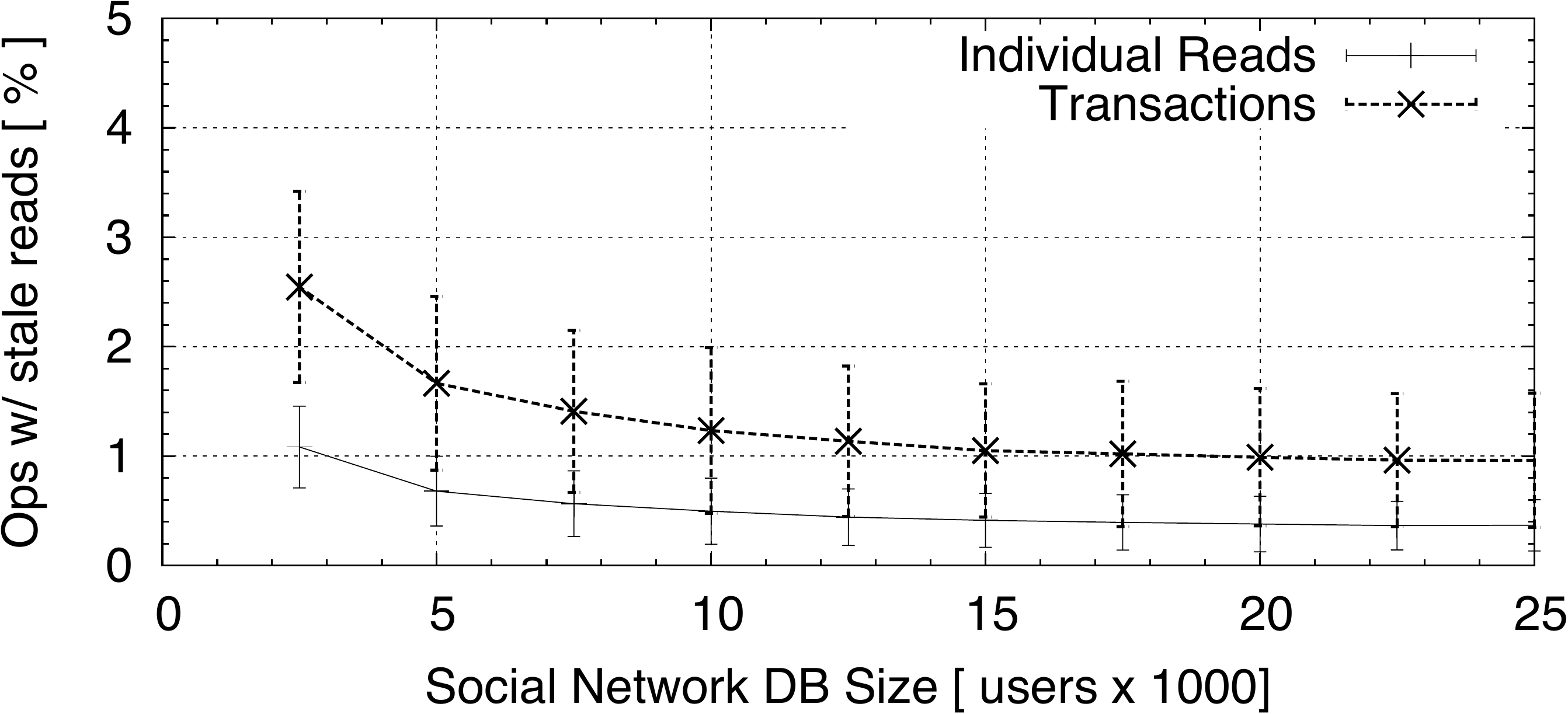}
\label{fig:stalereads}
\squeezeAboveCaption
\caption{Staleness of reads due to fault tolerance algorithm in \SwiftCloud as a function of contention (\SwiftSocial).}
\squeezeBelowCaption
\end{figure*}

\subsection{Staleness due to fault-tolerance}

We showed in Section~\ref{sec:eval:latency} that our approach to fault
tolerance minimises the latency perceived by end-users, compared to the
alternatives.
However, it slows down propagation of updates; our next experiment aims
to quantify by how much.
A read will be considered \emph{stale} if it returns a ($K$-durable)
version, and a more recent (non-$K$-durable) one exists and satisfies
\TransCausalPlus{}.
Preliminary work (not shown here)
showed that with the benchmarks used so far, the number
of stale reads is negligible.
The reason is that the window of vulnerability --- the time it takes for
a transaction to become $K$-durable --- is very small, approximately the
RTT to the closest DC.
We run the \SwiftSocial benchmark with 190~PlanetLab nodes spread across
Europe and five clients per node, connected to the Ireland DC and
replicated in the Oregon DC\@.
To further increase the probability of staleness, we make transactions
longer by setting the cache size to zero, requiring reads to contact a
DC, and commit to the farthest-away DC, with a RTT of around 170\,ms.

Figure~\ref{fig:stalereads} shows the occurrence of stale read
operations, and of transactions containing a stale read, for different
sizes of the database.
We have 950~concurrent clients; with 2,500 simulated users, at any time
approximately $40\%$ of the users are actively executing operations
concurrently.
Even in this case, stale reads and stale transactions remain under $1\%$
and $2.5\%$ respectively.
This number decreases as we increment the size of the database, as
expected.
This shows that even under high contention, accessing a slightly stale
snapshot has very little impact on the data read by transactions.

\begin{figure}[tp]
  \centering
  \includegraphics[width=0.58\columnwidth]{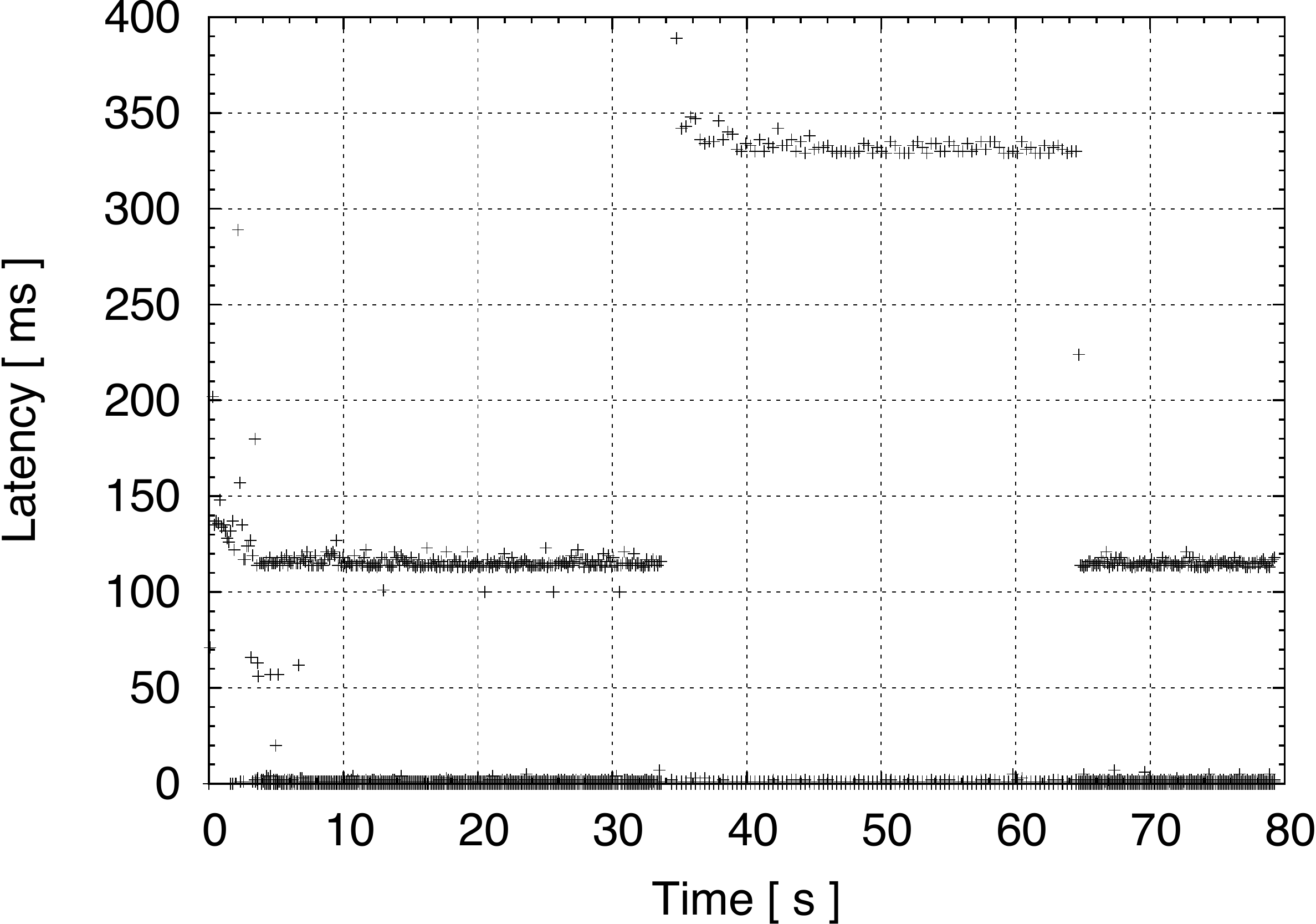}
\squeezeAboveCaption
  \caption{Latency for a single client switching data centres}
\squeezeBelowCaption
  \label{fig:handoff-plot}
\end{figure}
\subsection{Behaviour during faults}

Our final experiment studies the behaviour of \SwiftCloud{} when a DC
becomes disconnected or fails.
In this case,  clients fail-over to another DC\@.
The scatterplot in Figure~\ref{fig:handoff-plot} plots the latency of
transactions at an individual client as its scout switches DCs, while
running the \SwiftSocial{} benchmark.
Each dot represents the latency of an individual transaction.
Starting with a cold cache, latency quickly drops to near zero for most
transactions, those hitting in the cache, and to around 110\,ms for
those that perform remote reads due to cache misses.
Approximately 33\,s into the experiment, the scout is diverted to
another DC in a different continent.
The new latency pattern reflects the increased cost of cache misses, due
to the higher RTT to the DC, which also causes a visible drop in
throughput (sparser dots).
At 64\,s, the client switches back the initial data centre, and
performance smoothly recovers to the initial pattern.
Note that there are no significant gaps associated with switching,
showing that the protocol incurs negligible disruption to the client.

\section{Related work}	
\label{sec:soa}

Cloud storage systems provide a wide range of consistency models.
Some systems \cite{rep:pan:1693,pan:rep:1712} provide strong consistency
\cite{loo:syn:1468}, at the cost of unavailability when a replica is
unreachable (network partitions) \cite{rep:pan:1628}.
At the opposite end of the spectrum, some systems
\cite{syn:optim:rep:1433} provide only eventual consistency (EC), but allow
any replica to perform updates even when the network is partitioned.
Other systems' consistency models lie between these two extremes.

\textbf{Weak consistency:} 
Causal consistency strengthens EC with the guarantee that if a write is
observed, all previous writes are also observed. 
\citet{syn:rep:1677} show that, in the presence of partitions, this is
the strongest possible guarantee in an always-available, one-way
convergent system.
To cope with concurrent updates, Causal+ Consistency incorporate
mergeable data.
This is the model of COPS \cite{rep:syn:1662}, Eiger \cite{syn:rep:1708},
ChainReaction \cite{chainreaction} and Bolt-On \cite{bolt-on}.
These systems merge by last-writer-wins.
Some also support an application-provided merge function; for instance
Sporc \cite{rep:syn:1645} relies on operational transformation.

COPS and ChainReaction implement
read-only transactions that are non-interactive, i.e, the read set is
known from the beginning.
Eiger additionally supports non-interactive write-only transactions.
\SwiftCloudLong{} extends these works with interactive transactions,
integrated mergeable types support and support for DC failover.
A similar approach, including the study of session guarantees and atomicity,
was discussed by \citet{hatnotcap}.
\citet{par:lan:1665} and Orleans \cite{rep:syn:1663} also provide a model of
transactions for EC that
uses a branch-and-merge model with main revision, suitable for smaller databases.

Dynamo \cite{app:rep:optim:1606} and similar systems
\cite{rep:db:1688,rep:app:1682}
\cut{allow applications to control the level
of consistency by selecting the size of quorums.
In order to achieve high availability in the presence of failures,
these systems} ensure EC and per-key causality.
The timeline consistency of PNUTS \cite{rep:1685} and the snapshot
consistency of Megastore \cite{fic:syn:rep:1654} enforce a total order
on updates, but improve performance by allowing applications to read
stale data. 
Walter \cite{rep:syn:1661} and Gemini \cite{rep:syn:1690} support both
weak and strong consistency in the same system, for disjoint sets of
objects and of operations respectively.
Our support for non-mergeable transactions uses Gemini's
approach.

\cut{\textbf{Strong consistency:}
\comment[Marc]{This paragraph is pretty empty.  Cut it.}
Distributed database designs generally provide strong consistency
\cite{Ganymed,Tashkent,Granola,Thomson2012}.
These systems provide classical serialisable or snapshot-isolation
transactions with either full or partial replication.
We use some of the same techniques, e.g., identifying transactions
sequentially as in Calvin \cite{Thomson2012}, yet our approach is
fundamentally different.
Whereas these systems synchronise to detect concurrent and abort updates
(e.g., write-write conflicts), mergeable transactions do not need
coordination and automatically merge their effects.
}
\textbf{Concurrent updates:}
The last-writer-wins (LWW) rule \cite{db:rep:optim:1454} for for
managing concurrent updates selects between concurrent versions the one
with the highest timestamp
\cite{rep:syn:1662,syn:rep:1708,chainreaction,bolt-on,rep:db:1688}.
Depot \cite{rep:1712}, Dynamo \cite{app:rep:optim:1606} and CAC
\cite{syn:rep:1677} maintain all concurrent versions, letting the
application merge them somehow.

The theoretical basis for mergeable data is commutativity and lattice
theory.
Conflict-free Replicated Data Types (CRDTs)
\cite{rep:syn:1661,syn:rep:sh143}, proved to be mergeable using montonic
semi-lattice or commutativity, provide abstractions such as sets,
graphs, maps, counters and sequences.
$\text{Bloom}^L$ uses program analysis to check that a program's state
progresses monotonically in a semi-lattice, and if not inserts a
synchronisation point \cite{BloomL}; in comparison, our model does not
enforces determinism w.r.t.\ program input and avoids certain
synchronisation points, but puts more work on application programmer
to design transactions.

\SwiftCloud offers CRDTs because several useful abstractions are
available, richer yet subsuming LWW\@.
CRDTs were recently added to Riak \cite{rep:app:1682}
and Walter uses a set-like CRDT\@.
\SwiftCloud is the first mergeable-data system to support transactions
that span multiple CRDT types. 

\textbf{Fault-tolerance:}
With respect to tolerating DC faults, from the perspective of an
end-client, previous geo-replication systems fall into two categories.
Synchronous replication \cite{rep:pan:1693,Granola} can ensure that
clients observe a monotonic history in the presence of
DC faults, but at the cost of update latency.

Existing asynchronous replication systems ensure fault-tolerant
causal consistency only within the boundaries of the DC
\cite{rep:syn:1662,syn:rep:1708,rep:syn:1690,rep:syn:1661,bolt-on}.
Their clients do not keep sufficient information to ensure
causal consistency when a failure causes them to switch DCs.
These approaches trade low update latency for consistency or
availability.
To the best of our knowledge, \SwiftCloud is the first low-latency,
highly-available system that guarantees convergent causal consistency
with transactions all the way to resource-poor end clients.

\citet{bolt-on} observe that causal consistency can be
decomposed into separate safety and liveness components and that
presenting clients with stale versions can eliminate waiting for safety
dependencies.
We stretch this idea to the client that is not a full replica.

Session guarantee protocols \cite{rep:syn:1481} implement the safety
component of causal consistency.
\citet{k-durability} propose a protocol using $K$-durability.
This allows the client to change server, but their protocol is
synchronous and does not ensure exactly-once delivery.

Depot \cite{rep:1712} is the system most similar to \SwiftCloud.
Depot ensures causal consistency and high availability to clients, even
in the presence of server faults.
Clients can communicate directly with one another.
Depot is designed to tolerate Byzantine faults, a more
difficult class of faults than \SwiftCloud.
However it is not designed to scale to large numbers of clients, to co-locate
data with the user without placing a server in the user's machine, nor does it
support transactions.


\section{Conclusion} 
\label{sec:conclusion}

We  presented the design of \SwiftCloud, the first system that brings 
geo-replication to the client machine, providing a principled
approach for using client and data centre replicas.
\SwiftCloud allows applications to run transactions in the client machine,
for common operations that access a limited set of objects, or
in the DC, for transactions that require strong consistency 
or accessing a large number of objects.
Our evaluation shows that the latency and throughput benefit can be huge 
when compared with traditional cloud deployments for scenarios 
that exhibit good locality, a property verified in real 
workloads \cite{Benevenuto09Characterizing}. 

\SwiftCloud also proposes a novel client-assisted failover mechanism
that trades latency by a small increase in staleness. Our evaluation 
shows that our approach helps reducing latency while increasing stale reads 
by less than 1\%. 

Several aspects remain open for improvement.
Better caching heuristics, and support for transaction migration, would
help to avoid the high latency caused by successive cache misses.
Placing scouts at different levels of a hierarchy, in particular in Content
Delivery Network points of presence, might improve perceived latency even
more.
Finally, a better integration at the programming language level could help
address engineering concerns, such as data encapsulation across software stack.

\label{lastpage-before-bib}
{ \small \renewcommand{\baselinestretch}{0.8}
\newcommand{\textcommabelow}[1]{\c{#1}}
\bibliographystyle{abbrvnat}
\bibliography{predef,bib,shapiro-bib,local,marek}
}
\label{lastpage}
\end{document}